\documentclass[linenumbers, twocolumn]{aastex631}

\newcommand\kms{km s$^{-1}$}

\makeatletter
\DeclareRobustCommand{\SiII}{%
  \mbox{Si\check@mathfonts\fontsize\sf@size\z@\selectfont II}%
}
\makeatother

\makeatletter
\DeclareRobustCommand{\CII}{%
  \mbox{C\check@mathfonts\fontsize\sf@size\z@\selectfont II}%
}
\makeatother

\makeatletter
\DeclareRobustCommand{\HI}{%
  \mbox{H\check@mathfonts\fontsize\sf@size\z@\selectfont I}%
}
\makeatother

\DeclareRobustCommand{\ion}[2]{%
\relax\ifmmode
\ifx\testbx\f@series
{\mathbf{#1\,\mathsc{#2}}}\else
{\mathrm{#1\,\mathsc{#2}}}\fi
\else\textup{#1\,{\mdseries\textsc{#2}}}%
\fi}

\newcommand{\avg}[1]{\left<#1\right>}

\begin{document}

\title{A Simulated Galaxy Laboratory:\\Exploring the Observational Effects on UV Spectral Absorption Line Measurements}

\correspondingauthor{R. Michael Jennings}
\email{rjenni11@jhu.edu}

\author[0000-0002-3959-6572]{R. Michael Jennings}
\affiliation{Center for Astrophysical Sciences, Department of Physics \& Astronomy, Johns Hopkins University, Baltimore, MD 21218, USA}

\author[0000-0002-6586-4446]{Alaina Henry}
\affiliation{Center for Astrophysical Sciences, Department of Physics \& Astronomy, Johns Hopkins University, Baltimore, MD 21218, USA}
\affiliation{Space Telescope Science Institute; 3700 San Martin Drive, Baltimore, MD, 21218, USA}

\author[0000-0003-0595-9483]{Valentin Mauerhofer}
\affiliation{Kapteyn Astronomical Institute, University of Groningen, P.O Box 800, 9700 AV Groningen, The Netherlands}

\author{Timothy Heckman}
\affiliation{Center for Astrophysical Sciences, Department of Physics \& Astronomy, Johns Hopkins University, Baltimore, MD 21218, USA}

\author[0000-0002-9136-8876]{Claudia Scarlata} 
\affiliation{Minnesota Institute for Astrophysics, University of Minnesota, 116 Church Street SE, Minneapolis, MN 55455, USA}

\author[0000-0003-4166-2855]{Cody Carr} 
\affiliation{Center for Cosmology and Computational Astrophysics, Institute for Advanced Study in Physics, Zhejiang University, Hangzhou 310058, China} 
\affiliation{Institue of Astronomy, School of Physics, Zhejiang University, Hangzhou 310058, China}

\author[0000-0002-9217-7051]{Xinfeng Xu}
\affiliation{Department of Physics and Astronomy, Northwestern University,
2145 Sheridan Road, Evanston, IL, 60208, USA.}
\affiliation{Center for Interdisciplinary Exploration and Research in
Astrophysics (CIERA), Northwestern University, 1800 Sherman Avenue,
Evanston, IL, 60201, USA.}

\author[0009-0002-9932-4461]{Mason Huberty}
\affiliation{Minnesota Institute for Astrophysics, University of Minnesota, Minneapolis, MN 55455, USA}

\author[0000-0002-5659-4974]{Simon Gazagnes}
\affiliation{Department of Astronomy, The University of Texas at Austin, 2515 Speedway, Stop C1400, Austin, TX 78712-1205, USA}

\author[0000-0002-6790-5125]{Anne E. Jaskot}
\affiliation{Department of Astronomy, Williams College, Williamstown, MA 01267, USA}

\author[0000-0003-1609-7911]{Jeremy Blaizot} 
\affiliation{Univ Lyon, Univ Lyon1, ENS de Lyon, CNRS, Centre de Recherche Astrophysique de Lyon UMR5574, F-69230 Saint-Genis-Laval, France}

\author[0000-0002-2201-1865]{Anne Verhamme}
\affiliation{Univ Lyon, Univ Lyon1, ENS de Lyon, CNRS, Centre de Recherche Astrophysique de Lyon UMR5574, F-69230 Saint-Genis-Laval, France}
\affiliation{Observatoire de Gen\`eve, Universit\'e de Gen\`eve, 51 Ch. des Maillettes, 1290 Versoix, Switzerland }

\author[0000-0002-0159-2613]{Sophia R. Flury}
\affiliation{Department of Astronomy, University of Massachusetts Amherst, Amherst, MA 01002, USA}

\author[0000-0001-8419-3062]{Alberto Saldana-Lopez}
\affiliation{Observatoire de Gen\`eve, Universit\'e de Gen\`eve, 51 Ch. des Maillettes, 1290 Versoix, Switzerland}

\author[0000-0001-8587-218X]{Matthew J. Hayes}
\affiliation{Stockholm University, Department of Astronomy and Oskar Klein Centre for Cosmoparticle Physics, AlbaNova University Centre, SE-10691, Stockholm, Sweden}

\author[0000-0002-6849-5375]{Maxime Trebitsch}
\affiliation{Kapteyn Astronomical Institute, University of Groningen, P.O Box 800, 9700 AV Groningen, The Netherlands}



\begin{abstract}
Ultraviolet absorption line spectroscopy is a sensitive diagnostic for the properties of interstellar and circumgalactic gas.  Down-the-barrel observations, where the absorption is measured against the galaxy itself, are commonly used to study feedback from galactic outflows and to make predictions about the leakage of \HI{} ionizing photons into the intergalactic medium. Nonetheless, the interpretation of these observations is challenging and observational compromises are often made in terms of signal-to-noise, spectral resolution, or the use of stacking analyses. In this paper, we present a novel quantitative assessment of UV absorption line measurement techniques by using mock observations of a hydrodynamical simulation. We use a simulated galaxy to create 22,500 spectra in the commonly used \SiII{} lines while also modeling the signal-to-noise and spectral resolution of recent rest-frame UV galaxy surveys at both high and low redshifts. We show that the residual flux of absorption features is easily overestimated for single line measurements and for stacked spectra. Additionally, we explore the robustness of the partial covering model for estimating column densities from spectra and find under-predictions on average of 1.25 dex. We show that the under-prediction is likely caused by high-column-density sight-lines that are optically-thick to dust making them invisible in UV spectra.
\end{abstract}



\section{Introduction}
\label{sec:intro}
Over the last few decades, cosmological simulations have modeled the growth and evolution of galaxies, successfully reproducing many of the major scaling relations that characterize galaxies from early times to today \citep{Tremonti2004,Lilly2013}. A fundamental requirement in all simulations is the need for star formation feedback to produce the diverse populations of galaxies that we see today \citep{Silk2012}. By heating and ejecting gas, supernovae-driven feedback regulates star-formation, transports metals into the IGM, and creates a porous ISM that is essential for leaking \HI-ionizing photons and reionizing the IGM at early times \citep{Trebitsch2017,Barrow2020}. However, there is no clear agreement on the model for this feedback and a wide variety of implementations have been shown to reproduce many of the global properties of galaxies \citep{Vogelsberger2013,SomervilleDave15,Nelson2019}.

Direct observations of galaxies' outflowing gas can provide crucial input to models \citep{Chisholm2017,Chisholm2018}, potentially breaking degeneracies inherent in scaling-relation based constraints (e.g.\ the mass-metallicity relation). For this work, one of the most widely used tools is rest-frame ultraviolet (UV) spectroscopy, which covers a wealth of ISM absorption lines that appear blueshifted in the presence of an outflow. By measuring optical depths, covering fractions, column densities, and velocities in a diverse complement of ions, studies have estimated the rates of mass, energy, and momentum flow in galactic winds \citep{Heckman2015,Xu2022,Xu2023}. While early studies focused on $z\ga 1-2$, where the rest-frame UV can be observed with optical multi-object spectrographs \citep{Shapley2003,Weiner2009,Martin2012,Kornei2012,Erb2012}, the last decade has seen emphasis on the low redshift universe, following the installation of the Cosmic Origins Spectrograph (COS) on HST \citep{Heckman2011,Alexandroff2015,Henry2015, Chisholm2016a,Chisholm2016b,Chisholm2017_outflow}. 
Moreover, modeling UV absorption line profiles and their related resonant and fluorescent emission features has been proposed as a way to uncover the large scale geometry of the absorbing gas (\citealt{Carr2018,Carr2021} for idealized geometries, \citealt{Mauerhofer2021} for non-idealized geometries). 
    
Our physical understanding of galactic outflows is fundamentally linked to another key question: the escape of \HI-ionizing Lyman Continuum (LyC) photons, a phenomenon required to reionize the IGM at early times (see \citealt{Mesinger2016} for a review).  A clear understanding of both the large and small scale distribution of gas around galaxies is essential to predict the ionizing output of galaxies in the reionization epoch, where it cannot be measured.  Hence,  the same type of data-- UV absorption line spectra-- have been used to infer the amount of escaping LyC emission by determining the fraction of the stellar continuum that is uncovered by optically thick neutral hydrogen \citep{Heckman2011, Jones13, Borthakur14, Gazagnes2018, Chisholm2018, Saldana-Lopez2022}. 

Despite the efforts invested in UV spectroscopy for understanding these key measures of galaxy evolution, there are a number of challenges that observers have yet to overcome.  While it is generally understood that spectra must have good continuum signal-to-noise ($\ga 5-10$), and moderately high spectral resolution ($\la 100$ km s$^{-1}$ is desirable), these requirements are out-of-reach for the vast majority of galaxies at both low and high redshifts.  Ultraviolet instruments and detectors are only sufficiently sensitive for the most nearby, UV-luminous objects (e.g.\ \citealt{Berg2022, Xu2022}), while the apparent faintness of galaxies at $z>3$ has limited studies to rare highly magnified galaxies \citep{Quider2009, Quider2010, Jones13, Rigby2018, Chisholm2019}.  Compromises in signal-to-noise (S/N) or resolution are commonplace \citep{Shapley2003, Henry2015, Carr2018, Saldana-Lopez2022}, as in stacking to create composite spectra (e.g. \citealt{Weiner2009, Steidel2018}).  Critically, systematic assessments of these observational practices and their implications remain important areas of investigation. 

Further complicating matters, galaxies are complex three-dimensional bodies.  UV spectra are generally not spatially resolved (but see \citealt{James18, Keerthi2023}), and they represent only a single sight line towards a galaxy.  However, \cite{Gazagnes2023} showed that a {\it single} simulated $z=3-4.19$ galaxy, viewed from different orientations,  could reproduce the vast majority of \SiII{}1260 + \SiII{}$^*$1265 and \CII{}1334 + \CII{}$^*$1335 spectra in the COS Legacy Archive Spectroscopic SurveY (CLASSY; \citealt{Berg2022}), which spans four orders of magnitude in stellar mass around $z\sim 0.01$. Further, even a single sight-line towards a galaxy should exhibit a distribution of gas column densities and covering fractions, with variations in both position and line-of-sight projected velocity \citep{delaCruz2021}. Encoding all of this information into a single spectrum erases much of this detail, and it is fundamentally unclear if inferences made from spectra are representative. 

Mock observations of hydrodynamical simulations offer a means to test these observational practices in an environment where the ``truth'' is known, the spectral resolution is arbitrarily high, and signal-to-noise is only limited by Monte-Carlo photon statistics.  Therefore, in this paper, we carry out mock observations of a galaxy tracked from $z=4.19 $ to $z=3.00$ \citep{Mauerhofer2021} created using a radiation hydrodynamical code \citep{Rosdahl2015}. The simulations are post-processed with Monte-Carlo radiative transfer \citep{Michel-Dansac2020, Mauerhofer2021} to produce UV ISM absorption features, thereby producing realistic spectra that account for resonant scattering and emission. By modeling the observational effects inherent in recent rest-frame UV galaxy surveys, we can assess the reliability of spectral stacking methods and the accuracy of column density estimates using the apparent optical depth method \citep{Savage1991} with a partial covering model (PCM; \citealt{Rupke2005}) in data with a range of quality, similar to some of the important UV spectroscopic surveys in the literature today. The purpose of this paper is to use the simulated galaxy as a ground truth for testing observation effects on spectra and \textit{not} as a ground truth for real galaxies. Therefore, throughout the paper we avoid any direct comparisons between the simulation and the survey observations.

This paper is organized as follows:  
We introduce the simulation and our methods for creating the mock observations in section~\ref{sec:data}. We then introduce the calculations for spectral line measurements in section~\ref{sec:observational effects}. Next, we explore the effects of spectral stacking on line measurements in section~\ref{sec:stacking effects}. We then discuss the implications of our findings on escape fraction estimations in section~\ref{sec:wrv_discussion}. Testing of the partial covering model is presented in section~\ref{sec:column densities}. Lastly, we summarize our findings in section~\ref{sec:summary}.

\section{Data}\label{sec:data}

\begin{deluxetable*}{cccccccccc}
\tablecaption{Representative data used to generate the mock survey spectra.}
\colnumbers 
\tablehead{
\colhead{Survey} & \colhead{Ref.} & \colhead{$\log{M_{*}}$} & \colhead{$\log{\text{SFR}}$} & \colhead{$Z_{\mathrm{gas}}$} & \colhead{$z$} & \colhead{$\sigma_{\text{FWHM}}$} &  \colhead{$\Delta \lambda$}  & \colhead{S/N} & {$\sim W_{\text{lim}}$} \\
\colhead{} & \colhead{} & \colhead{($M_{\sun}$)} & \colhead{($M_{\sun}\,\text{yr}^{-1}$)} & \colhead{($Z_{\sun}$)} & \colhead{} & \colhead{(km s$^{-1}$)} & \colhead{(\AA, km s$^{-1}$)} & \colhead{} & \colhead{(\AA)}}
\startdata
Simulation & a & 8.85--9.36 & 0.13--0.60 & 0.21--0.42 & 3.00--4.19 & - & 0.04, 10 & 20 & 0.025, 0.049, 0.074 \\
LzLCS G140L & b & $8.95^{8.39}_{9.98}$ & $1.17^{0.53}_{1.89}$ & $0.30^{0.13}_{0.48}$ & $0.31^{0.24}_{0.36}$ & 300 & 0.43, 103 & 4 & 0.39, 0.78, 1.17 \\
CLASSY & c & $8.12^{6.65}_{9.58}$ & $0.39^{-1.44}_{1.46}$ & $0.26^{0.075}_{0.50}$ & $0.022^{0.0026}_{0.13}$ & 65 & 0.18, 43 & 11.2 & 0.09, 0.18, 0.27 \\
VANDELS & d & $9.68^{9.06}_{10.70}$ & $1.12^{0.26}_{1.79}$ & - & $3.33^{1.35}_{4.27}$ & 461 & 0.64, 152 & 10 & 0.19, 0.38, 0.57 \\
\enddata
\vspace{-0.1cm}
\tablecomments{(1) Name of survey (with HST/COS grating if applicable). (2) Survey reference(s): a = \citet{Mauerhofer2021}, b = \citet{Flury2022a}, c = \citet{Berg2022,Xu2022}, d = \citet{Garilli2021}. (3) Stellar mass. Simulation reports full range measured within 1 virial radius while other surveys report median with 10th and 90th percentiles. (4) Star formation rate. Simulation reports full range measured within 1 virial radius and averaged over the last 100 Myr. (5) Gas phase metallicity. Simulation reports full range measured within 1/10 virial radius. VANDELS survey has no reported value in reference. (6) Representative redshift. (7) Spectral resolution (i.e. full width at half maximum of Gaussian kernel used in convolution to downgrade simulation spectra.) (8) Dispersion binning. (9) Continuum signal-to-noise ratio (10) Limiting equivalent width measurable to 1$\sigma$, $2\sigma$, $3\sigma$.}
\label{tab:surveys}
\end{deluxetable*}

\subsection{Simulation}\label{sec:simulation}
In this paper we use mock galaxy spectra derived from a cosmological zoom-in simulation. The details of the simulation can be found in \citet{Mauerhofer2021} with updates described in \citet{Gazagnes2023} and \citet{Blaizot2023}. We review the necessary information here. The simulation is run using {\tt RAMSES-RT} \citep{Teyssier2002,Rosdahl2013,Rosdahl2015}, an adaptive mesh refinement code including radiative transfer. {\tt MUSIC} \citep{Hahn2011} is used to generate the initial conditions to produce the desired galaxy with stellar mass $M_{*} \sim 10^9 M_{\sun}$ with a maximum cell resolution of 14 pc at $z = 3$. Gas cooling is allowed down to 15K, allowing for a naturally fragmenting ISM and molecular star forming regions where the subgrid star formation model of \citet{Kimm2017} is employed. Across the 75 time snapshots the stellar mass and gas metallicity of the galaxy steadily increase to values of  $\log{(M_{*}/M_{\sun})} = 9.36$ and $Z_{\text{gas}}/Z_{\sun} = 0.42$, with the star formation (averaged over 100 Myr) oscillating around $\log{(\text{SFR}/M_{\sun}\,\text{yr}^{-1})} = 0.44$. The full galaxy property ranges are listed in Table~\ref{tab:surveys}.

\subsection{Simulated Spectra}\label{sec:simulated spectra}
The simulation outputs are post-processed as described in \citet{Mauerhofer2021}, thereby generating mock spectra of metallic ions of interest. The densities of metallic species are calculated using the hydrogen and helium densities and the metal mass fractions $Z$ which are traced by the simulation. Solar abundance ratios are assumed \citep{Grevesse2010}. Ionization fractions are then computed using the chemical network evolution code {\tt KROME} \citep{Grassi2014} with the ionization processes of recombination, collisional ionization, and photoionization included. Ion populations between the fine-structure split ground states are calculated using PyNeb and these multiple lines are then followed in the subsequent radiative transfer post-processing using RASCAS \citep{Michel-Dansac2020}. 

With RASCAS, photon packets are launched from stellar particles based off of BPASS spectral energy distributions with a given direction and frequency and propagated through the simulation grid such that the spectra have a stellar continuum. Photon packets have a probability of interaction in each grid cell according to the optical depth of each of several channels: one of each ionization levels and one dust channel. Photon packets are allowed to be absorbed or scatter upon interaction with dust. The optical depth of dust is produced as a post-processing effect and uses a pseudo-density model that is proportional to the density of hydrogen \citep[see][]{Mauerhofer2021} and assumes an albedo value of 0.32 \citep{Li2001}. Upon absorption by any ion channel the photon packet will be re-emitted in an isotropic manner either via resonant scattering or the fluorescent channel if absorbed by the appropriate transition. The radiative transfer calculation of a photon packet ends when absorbed by dust or upon exit from the galaxy virial radius. For this reason, photon escape fractions of the simulation are calculated by counting the number of photons leaving the virial radius. 

Our fiducial mock observations--identical to those presented in \citet{Gazagnes2023}--are made from a total of 300 directions for each of the 75 snapshots of the galaxy with chosen spectral resolution and aperture radius, resulting in a total of 22,500 unique sightlines. The spectral resolution is chosen to be 10 \kms\ with an aperture of 1" in diameter, which corresponds to a physical size of $\sim7-8$kpc at the redshifts of the simulation outputs. This aperture size allows for the entirety of the ISM as well as scattered light from the CGM. Although this aperture size is much larger than those used in our comparison surveys, \citet{Gazagnes2023} showed that the main effect of reduced aperture size is the loss of extended fluorescent emission in the resultant spectra. Our focus on absorption line analysis in this paper is therefore not greatly effected by the large choice in aperture.

\subsection{Mock Survey Observations}\label{sec:mock observations}

\begin{figure*}
    \centering
    \includegraphics[width=\textwidth]{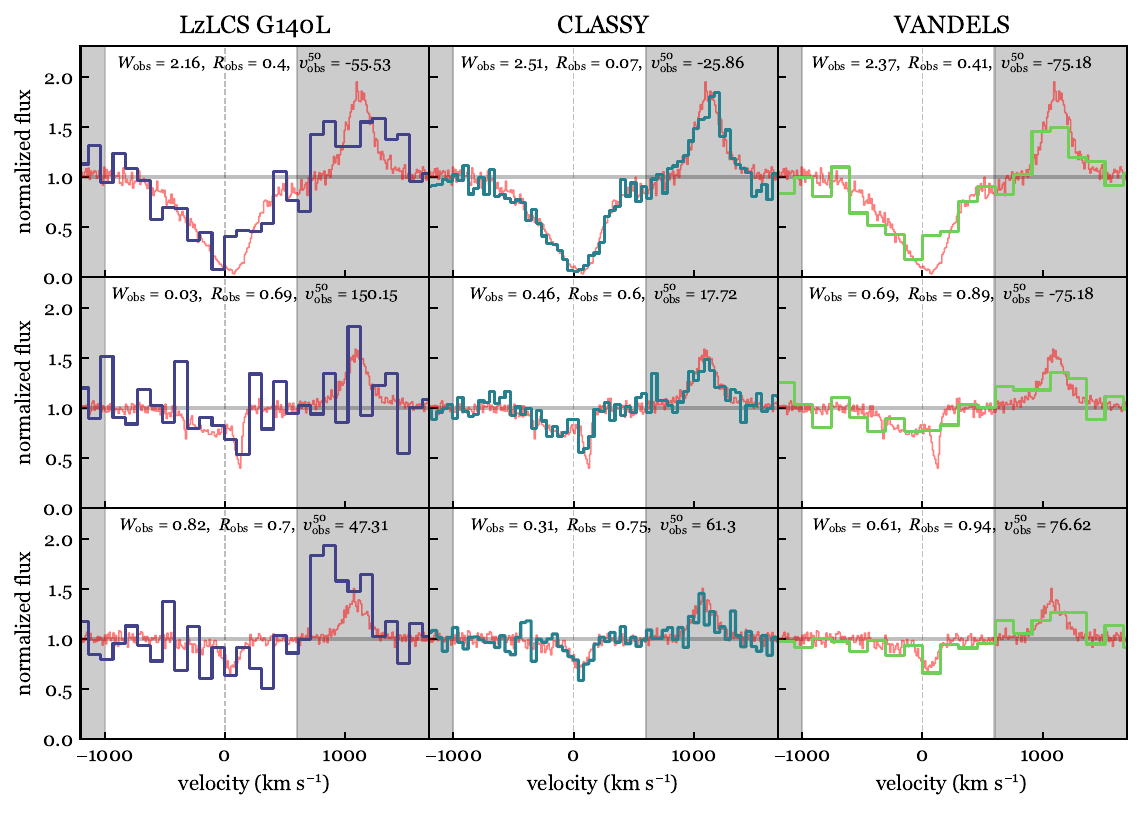}
    \caption{Example spectral observations of the simulated galaxy for \SiII{} $\lambda 1260$\AA. Each row shows a different observed sightline of the galaxy with decreasing line strength from top to bottom with $W_{\text{sim}}=2.4,0.6,0.3$\AA; $R_{\text{sim}}=0.04,0.44,0.7$; $v^{50}_{\text{sim}}=-2.47,-62.26,24.14$. Each column shows the survey data the mock spectra represent. The true simulation spectra are shown in red with the mock observation spectra plotted on top. The un-shaded region shows the integration range for equivalent width measurements.}
    \label{fig:spec_examples}
\end{figure*}

In order to investigate observational effects, we create mock observations designed to match those from the Low-Redshift Lyman Continuum Survey (LzLCS) \citep{Flury2022a,Flury2022b}, the COS Legacy Archive Spectroscopic Survey (CLASSY) \citep{Berg2022,James2022} and VANDELS  \citep{Garilli2021}. The LzLCS and CLASSY samples consist of 66 and 45 galaxies observed in the far-ultraviolet using the Cosmic Origins Spectrograph (COS) on the Hubble Space Telescope (HST). The VANDELS survey, a European Southern Observatory Public Spectroscopic Survey using the VLT/VIMOS instrument, provides spectra for 2087 galaxies with redshifts between $z \sim 1-6$ at medium resolution.   

A majority of the LzLCS sample were observed using the low resolution G140L grating and so we downgrade the simulated spectra resolution to match the resolution of the data. For observations of spatially resolved galaxies with COS, the line spread function is determined primarily by the size of the galaxy in the dispersion direction, resulting in a resolution that is degraded relative to that of a point source.  Therefore, we adopt the effective spectral resolution quoted in the relevant galaxy studies. To downgrade the simulated spectra we convolve with a Gaussian kernel with a full width at half maximum of $\sigma_{\text{FWHM}}$ to match the resolution of the survey under consideration. The typical resolution of the LzLCS sample observed with the G140L grating is $R \equiv \lambda/\sigma_{\text{FWHM}} \sim 1000$ at the observed wavelength of $1100$ \AA\ \citep{Izotov2021,Flury2022a} with observed spectral dispersion of $\Delta \lambda = 0.5621$ \AA. This corresponds to a spectral resolution of $\sigma_{\text{FWHM}} \sim 300$ \kms\ and rest-frame binning of $\Delta \lambda \approx 0.43$ \AA\ at $z=0.3$. The highest resolution co-added spectra of the CLASSY sample use the G130M grating with a median effective spectral resolution of the spatially resolved galaxies being $\sigma_{\text{FWHM}} \sim 65$ \kms\ (see Table 3 of \citealt{Berg2022}) and an average S/N$\sim 6.4$ (see Table 1 of \citealt{Berg2022}). We use the binning choice of \citet{Xu2022}, a rebinning of 3:1 to $\Delta \lambda \approx 0.18$ \AA\ to increase the average signal-to-noise per pixel by $\sqrt{3}$ to S/N $\sim 11.2$. For the VANDELS survey we adopt a conservative representative S/N of 10 where virtually all star forming galaxies of the survey exceed this value. These representative values used to mock each survey are shown in Table~\ref{tab:surveys}.

Next, we match the noise of the survey spectra. To do this we add Gaussian random noise to the simulated spectra's normalized flux with a standard deviation of  $\sigma_{\text{norm}} \approx 1/(\text{S/N})$. Because the lower resolution surveys are dominated by background and dark-current noise we then add Gaussian random noise with standard deviation of $\sigma_{\text{norm}}$ to the normalized flux in each wavelength bin. For the CLASSY survey with brighter sources we add Gaussian random noise with standard deviation of $\sigma_{\text{norm}} \sqrt{F}$, with $F$ being the normalized flux value. The square-root coefficient mimics the nature of the flux uncertainty due to Poisson statistics on the photon counts.

To capture the spread of possible measurements from the same sightline with different realizations of noise, we sample the noise 500 times. This gives us a total of $(75\, \text{outputs})\times (300\, \text{directions})\times (500\, \text{noise samples}) = 11,250,000$ spectral samples. This allows us to explore observation and methodology effects from a statistical advantage.

\section{Observational Effects}\label{sec:observational effects}

\begin{figure*}
\centering
\includegraphics[width=\textwidth]{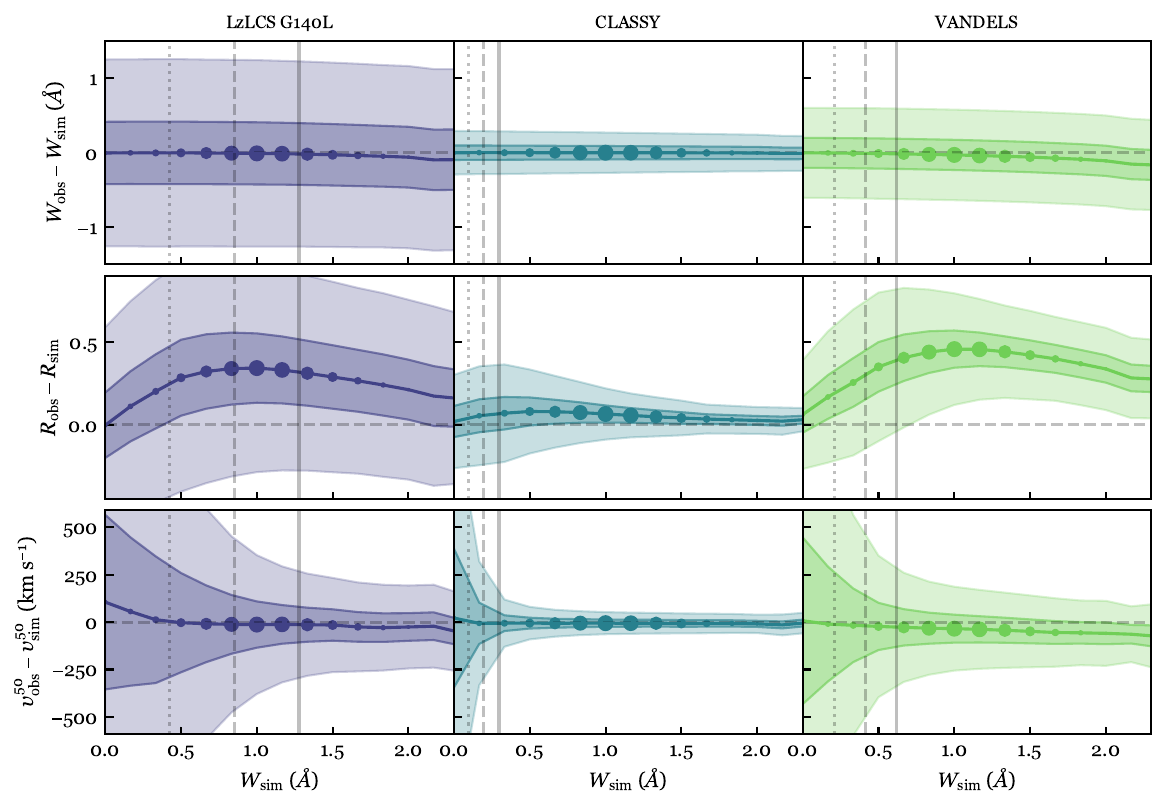}
\caption{\SiII{} $\lambda$1260 differences of equivalent width, residual flux, and 50\% velocity for the three mock surveys. Each difference is plotted as a function of ``true'' equivalent width. The solid lines show the median, and the shaded regions show the $1\sigma$ and $3\sigma$ range respectively. The size of the markers is proportional to the count of sightlines within each $W_{\text{sim}}$ bin. The vertical dotted, dashed, and solid lines for each survey column represent the respective $W_{\text{lim}}$ value at $1\sigma,2\sigma,3\sigma$.
\label{fig:all_residuals}}
\end{figure*}

\begin{figure*}
\centering
\includegraphics[width=\textwidth]{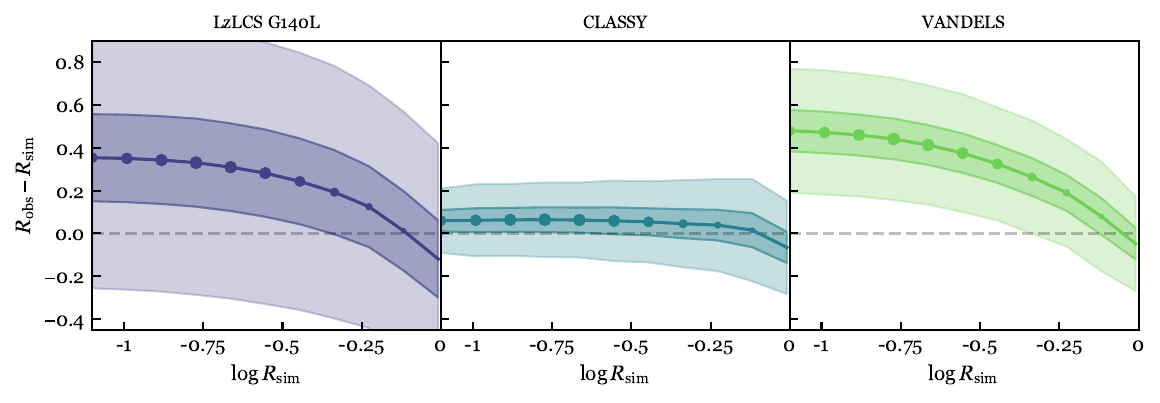}
\caption{Same as the middle row of Figure~\ref{fig:all_residuals} but plotted as a function of ``true'' residual flux.
\label{fig:all_residuals_R}}
\end{figure*}

We first focus on the observational effects on individual spectra. In Figure~\ref{fig:spec_examples} we show example \SiII{} $\lambda 1260$  observations of the simulated galaxy from three different sightlines with varying degree of line strength. The processed spectra mock LzLCS sample observations using the HST/COS G140L grating, the high resolution CLASSY observations and the VANDELS survey. Two obvious effects come from downgrading the resolution and the addition of noise. As expected, the 300 \kms and 461\kms convolutions to match the LzLCS G140L and VANDELS surveys, significantly diminishes and widens the line features while the addition of noise can greatly obfuscate the line position and depth. Both of these procedures affect our measurements of equivalent width, residual flux and velocity in various ways as we discuss below.

\subsection{Equivalent Width}\label{sec:equivalent width}
The equivalent width, $W$, of an absorption feature is proportional to the total missing power that has been removed along the line of sight by the absorbing material and is conserved even for unresolved lines. It is a common assessment of absorption and emission line strength. Therefore, we use equivalent width to validate our procedures and understand the effects of noise on line measurements with varying S/N.

We use the conventional definition to calculate the equivalent width of a line for a given sightline $j$ according to:
\begin{equation} \label{eq:W}
W_j = \sum_i^{x} (1-F_{ij})\,\Delta \lambda,
\end{equation}
where $F_i$ is the normalized flux in wavelength bin $i$, equal to the observed flux divided by the continuum flux. The continuum flux is a linear fit in a feature-free range around each line before observational effects are applied. In this definition, absorption is positive while emission is negative. The appropriate integration range, $x$, is more straightforward to determine on high resolution, high S/N spectra like the unprocessed simulation spectra. In this ideal case, the minimum of the absorption line is found and integration is continued separately in the blue and red directions until the continuum flux level is reached. However, for low resolution, low S/N spectra where the boundaries of the absorption feature can be hard to determine, a fixed integration boundary can be used. For consistent comparison, we take the approach of \citet{Saldana-Lopez2022} for all survey data and use a fixed value of $-1000\,\text{\kms} \leq x \leq +600\,\text{\kms}$ (chosen to include the absorption feature while excluding the emission feature for the strongest lines) shown by the unshaded regions in Figure~\ref{fig:spec_examples}. 

Our measurements of equivalent width are expected to be affected most by the addition of noise to the spectrum. The convolution and binning procedures should conserve the integrated area of the line preserving the equivalent width within a small percentage. However, the addition of Gaussian noise on top of the convolved and binned line can introduce deviations from the ``true'' simulation equivalent width. Therefore, we also quantify the fidelity of an absorption feature observation for each mock survey against the raw RASCAS output spectra. We use reported limiting equivalent widths $W_{\text{lim}}$ for HST/COS data \citep{Ghavamian2010,Keeney2012} and adapt the expression for our specific approach:
\begin{equation} \label{eq:W_lim}
W_{\text{lim}} \approx N_{\sigma} \sigma_{\text{noise}} \sqrt{x \Delta \lambda},
\end{equation}
where $N_{\sigma}$ is the desired significance level of the limiting measurement (i.e. 1$\sigma$, $2\sigma$, $3\sigma$, etc.).

To build statistics of the observational effects we use the 500 realizations of the mock noise on each sightline spectrum and calculate the equivalent width, giving a total of 11,250,000 measurements. The results of these calculations are compared to, and binned by, the ``true'' equivalent width $W_{\text{sim}}$ (measured on the unprocessed simulation spectra) and shown in the top row of Figure~\ref{fig:all_residuals}. As expected, the median of the differences between the observed equivalent width, $W_{\text{obs}}$, and the simulation equivalent width, $W_{\text{sim}}$, straddle a line near zero showing that on average the observed equivalent width matches the true equivalent width (in other words, the Gaussian random noise is averaged out). The spread in the differences are shown in the shaded regions signifying the 1$\sigma$ and 3$\sigma$ ranges. The spread in differences for each mock survey decreases with increasing S/N with the LzLCS G140L, CLASSY and VANDELS mock spectra measuring 68\% (99.7\%) of the equivalent width differences within $0.4,0.04,0.1$\AA{} ($1.2,0.1,0.7$\AA) respectively.

Figure \ref{fig:all_residuals} does reveal a subtle trend in the recovery of the  equivalent width for the low-resolution LzLCS and VANDELS mock observations.  $W_{\text{obs}}$ is underestimated by up to 0.1 \AA\ for true equivalent widths $\gtrsim 1.5$\AA.  As this trend is not apparent in the high resolution mock observations, we attribute it to the fixed integration window and blending of fluorescent emission from  \SiII{}* $\lambda$1265, which is separated from the absorption by $\sim 1000$ \kms.  Since we have calculated $W_{\text{obs}}$ using a fixed velocity range of -1000 \kms $\le v \le$ 600 \kms, some \SiII{}* emission likely bleeds into this range when this feature is strong and the spectral resolution is low. The trend with equivalent width in Figure \ref{fig:all_residuals} is consistent with this scenario, although the amplitude of the effect is smaller than the typical measurement uncertainty for individual lines.

While the amplitude of equivalent width bias due to blending with emission is small for the LzLCS and VANDELS surveys and non-existent for the higher resolution case of CLASSY in Figure~\ref{fig:all_residuals}, the true impact of this effect will vary.  In particular, the significance of this blending is dependant upon the instrumental resolution, wavelength separation of the absorption and emission features, and the strength of the line. For example, the \ion{C}{II} $\lambda 1334$\AA{} absorption line and its $\lambda 1335$\AA{} fluorescent emission line are separated by $\sim 250$\kms\ and can undergo significant blending compared to the \SiII{} $\lambda 1260$\AA{} line ($\sim 1000$\kms\ absorption and fluorescent emission separation).  On the other hand, as noted in \citet{Gazagnes2023}, real observations like the LzLCS and CLASSY will likely experience less contamination of their \SiII{} absorption due to fluorescent emission because of the smaller aperture size. Still, observational investigations should consider whether their measurements may be impacted by this effect, in addition to the well-established ``infilling'' of the absorption and P-Cygni emission \citep{Prochaska2011, Scarlata2015}.    

\begin{figure}
\centering
\includegraphics[width=\columnwidth]{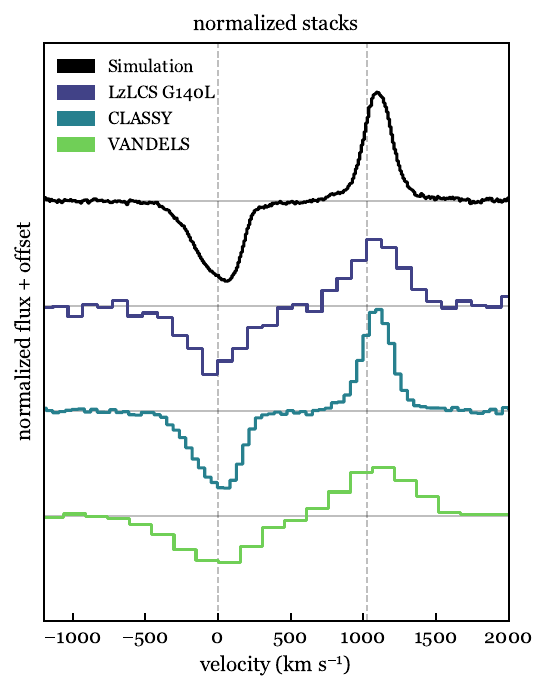}
\caption{Example stacks of normalized \SiII{} $\lambda 1260$ spectra for each survey. Each stack consists of the same 30 sightlines.
\label{fig:stack_examples}}
\end{figure}

\subsection{Residual Flux}\label{sec:residual flux}
The residual flux, $R$, of an absorption feature is the fraction of the total flux at the deepest part of the line. This line depth value can be used as an indication of the amount of column of gas present for optically thin lines. For saturated lines it is often interpreted as a gas covering fraction of the source \citep{Vasei2016,Reddy2016,Gazagnes2018,Steidel2018,Saldana-Lopez2022} which is then commonly used in estimations of the Lyman continuum escape fraction \citep{Heckman2011,Borthakur14}. As a simple and powerful gas diagnostic, accurate residual flux measurements are highly sought after. 

The residual flux is straightforward to determine from the high resolution, high signal-to-noise observations but more care must be taken when measuring $R$ from the lower resolution spectra. We take the approach of \citet{Saldana-Lopez2021} for all survey data and find the minimum flux of the line and then calculate the median flux over a fixed velocity range (+1 velocity bin on either side of minimum flux pixel) around the minimum to avoid biasing towards minimum depth bins generated from noise.

Whereas equivalent width is preserved under convolution and re-binning, it is well understood that residual flux is not. The convolution process when downgrading the spectral resolution redistributes the area in the deepest parts of line features into the wings creating a shallower and wider line profile. This, along with re-binning to lower dispersion, smooths and homogenizes fine details in the line profile. This effect results in the overestimation of $R_{\text{obs}}$ in the middle row of Figure~\ref{fig:all_residuals} and \ref{fig:all_residuals_R}. The narrowest and deepest absorption lines (low residual flux) in the intermediate equivalent width range are the most largely affected by the convolution and re-binning process while the shallowest and broadest lines are mostly unchanged. This effect is highlighted when comparing low and high S/N surveys. When convolving with a higher resolution line spread function, more narrow details are preserved and the line undergoes less redistribution of area into the wings. However, even for the higher resolution and S/N surveys, residual flux can be overestimated by up to $\sim 0.2$ at $3\sigma$ with a median systematic overestimation of $\sim 0.05$; a small but noticeable trend. In the most severe case, for the LzLCS, the detectable range of equivalent widths shows that $R$ is overestimated by $\sim 0.3-0.4$, and that this value is relatively flat with equivalent width. These results confirms the known effect of overestimating residual flux for underesolved lines \citep[as also presented in][]{Saldana-Lopez2023}.

\subsection{Velocity}\label{sec:velocity}
Velocity measurements on blueshifted UV absorption lines are key for quantifying outflow kinematics, mass loss and energetics for galaxies (e.g., \citealt{Xu2022}). For measuring the velocity of absorption lines we take a common approach \citep{Chisholm2017,Gazagnes2023} and choose to define a value $v^{50}$ which represents the velocity at which 50\% of the equivalent width is reached starting from the redward integration boundary.

Due to our choice in definition, the velocity measurements are limited by and affected in the same manner as the equivalent width measurements. However, as the equivalent width measurements showed the mean residual continuing to straddle zero with decreasing true equivalent width, the velocity mean residuals show a clear overestimation at smaller equivalent width measurements. When attempting to extract $v^{50}$ values on weak line features, noise becomes the dominating factor in skewing the results away from the true value. This emphasises the need to determine if a line feature is a non-detection (see where the vertical lines representing $W_{\text{lim}}(1\sigma,2\sigma,3\sigma)$ overlap velocity residuals in Figure~\ref{fig:all_residuals}). However, the measurements of $v^{50}$ on average are quite representative of the true values for lines with true equivalent widths detected above the $3\sigma$ significance. Nonetheless, in the LzLCS G140L spectra, the velocity uncertainties, as indicated by the purple shaded regions, are large ($\sim 300$ \kms at $W_{\mathrm{sim}} \sim 1.5$\AA) compared to a typical outflow velocity measurement.

\section{Stacking Effects}\label{sec:stacking effects}

\begin{figure*}
\centering
\includegraphics[width=\textwidth]{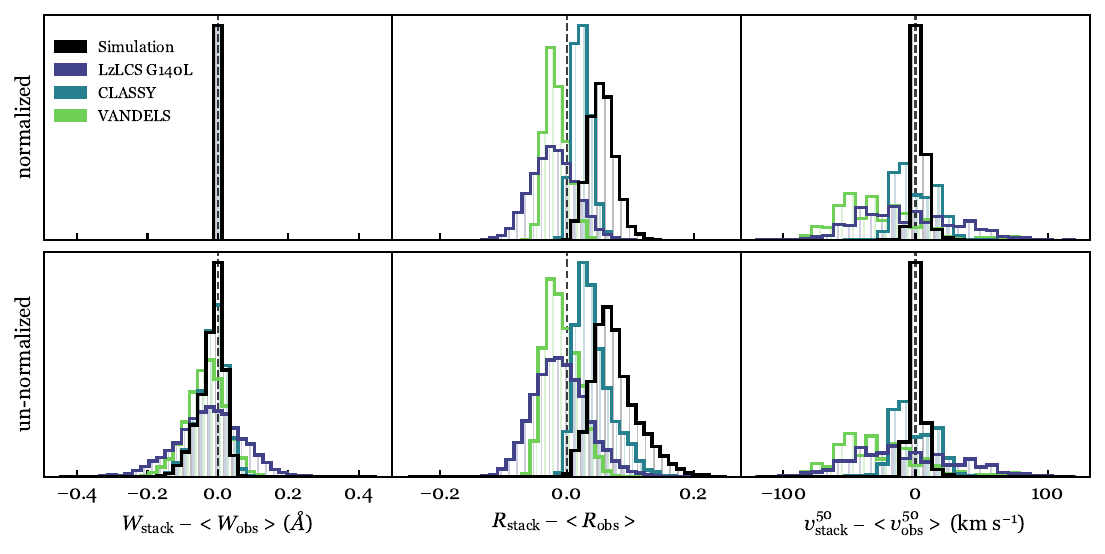}
\caption{Measurement residuals on stacked spectra for each survey. Our three line measurements are shown in each column starting with equivalent width on the left, residual flux, then 50\% velocity on the right. Two methods of stacking are shown with normalized flux average on the top row and a un-normalized average on the bottom. The un-normalized stacked measurement is compared to the continuum flux weighted average of the constituent $W$, $R$, and $v^{50}$.
\label{fig:all_stacks_2methods}}
\end{figure*}

Stacking, the process of averaging the signal of multiple spectra, has commonly been used \citep{Shapley2003,Jones2012,Trainor2015,Du2018,Pahl2020} to improve the S/N of faint galaxy observations in hopes of extracting representative information of the sample set. Despite its use, stacking methods have not been thoroughly investigated to understand how measurements should be interpreted \citep[see][for a recent investigation]{Saldana-Lopez2023}. We now highlight the effects of stacking on line measurements. We create a single stacked spectrum of the \SiII{} 1260\AA{} line by averaging the flux of $N = 30$ sightlines sampled from an individual galaxy time output. This way each spectrum in the stack has identical global galactic properties and the only variables contributing to stacking effects are sightline orientation. We also explored stacking across time stamps, thereby emulating the effect of stacking the spectra of different galaxies with similar properties.  Nonetheless, owing to the limited dynamical range of mass, SFR, and metallicity (see~\ref{sec:simulation}), we see negligible variations on results (rather, sightline-to-sightline variations dominate output-to-output variations). We choose 30 sightlines to roughly match the number of galaxies in LzLCS and CLASSY \citep[e.g.][ stacks 31 CLASSY galaxies to estimate sulfur column densities]{Xu2022}.

We create stacks of the fiducial simulated spectra as well as the mock survey spectra in order to investigate how observational effects change the outcomes of stacked measurements. In addition to the LzLCS and CLASSY cases, we model the observational effects of ground-based spectroscopy of $z\sim3$ galaxies from VANDELS, which has resolution and signal-to-noise characteristic of rest-frame UV spectroscopic surveys carried out on 8-10 meter class telescopes with multi-object optical spectrographs \citep{Shapley2003,Steidel2010,Jones2012,Trainor2015,Steidel2018,Du2018,Pahl2020}. Results from such ground-based surveys rely heavily on stacking, so their inclusion here is especially pertinent.

We also investigate two methods for spectral stacking: normalized and un-normalized flux. For the normalized stacks, we take the average of normalized fluxes within each wavelength bin. This first approach makes comparing the individual line measurements to the stacked measurement straightforward. For the un-normalized stacks, we take average of un-normalized fluxes within each wavelength bin. In this case, we calculate the representative $\avg{W_{\text{obs}}}$, $\avg{R_{\text{obs}}}$, and $\avg{v^{50}_{\text{obs}}}$ by weighting the individual line measurement according to the continuum flux for each spectrum. We note that these weighted averages are generally not recoverable in observations where stacking is deemed necessary for absorption line measurements.

For each survey using a given stacking method we generate 7,500 stacks: 100 stacks made of random sets of 30 sightlines for each of 75 galaxy outputs. Example stacks of both methods for each mock survey are shown in Figure~\ref{fig:stack_examples}. For each of the 7,500 stacks, we measure the equivalent width, residual flux, and velocity, and the same quantities from the 30 constituent sightlines. We calculate the difference between the stack measurement and the mean\footnote{Additionally, we tested median stacks and found marginal differences.} (or flux-weighted mean) quantities derived from the individual mock spectra with included observational effects--resolution, binning, and noise--and show the distribution of these differences for the 7,500 stacks in Figure~\ref{fig:all_stacks_2methods}. This shows how closely related the line measurement on the stack is to the average value of the constituent samples.

\subsection{Equivalent Width}
The first column on the left shows the difference in the equivalent width measurement $\Delta W \equiv W_{\text{stack}}-\avg{W_{\text{obs}}}$ with, $\avg{W_{\text{obs}}}$ being the average equivalent width of the mock observations (for the ``un-normalized'' bottom row, each spectrum is weighted by its continuum flux value) and $W_{\text{stack}}$ being the equivalent width of the stacked spectrum. The equivalent width measurements for the normalized stacks is found to be $\Delta W \equiv W_{\text{stack}}-\avg{W_{\text{obs}}} = 0$ for all surveys. This is expected to be identically zero because $\sum_i^x\sum_j^N F_{ij} = \sum_j^N\sum_i^x F_{ij}$ due to the commutative property for finite sums. This shows that an equivalent width measurement on a pre-normalized unweighted stacked spectrum perfectly relays the average of the equivalent widths of the individual spectra that make up the stack. However this is not true for the un-normalized stack. Our choice of weighting by average continuum flux introduces uncertainty on measuring the continuum. However, all survey distributions for the residual peak around zero. This shows that both methods for stacking quite robustly give equivalent width measurements that represent the average of the individual spectra with a preference for the normalized method. Furthermore, because the flux weighted averages of the individual measurements are usually unknown, we suggest using the normalized stacking method.

\subsection{Residual Flux}
The middle column of Figure~\ref{fig:all_stacks_2methods} shows the difference in the residual flux measurements. The measurement $\Delta R \equiv R_{\text{stack}}-\avg{R_{\text{obs}}}$ shows the difference between the average of the individual residual flux measurements of the constituent spectra and the single residual flux measurement on the stacked spectrum. Again, $\avg{R}$ is calculated from the mean of the measurements on spectra that include observational resolution and signal-to-noise. Hence,  this overestimation of $R$ is in addition to the effects on individual spectra that we demonstrated in Section~\ref{sec:observational effects}. Unlike for $\Delta W$, we find that $\Delta R \equiv R_{\text{stack}}-\avg{R} \neq 0$ for all surveys, including the native simulation spectra. All distributions peak off of zero, meaning the residual flux of a stacked spectrum will tend to be higher or lower than the average residual fluxes. This is particularly of interest, as covering fraction and escape fraction of LyC photons are often inferred by residual flux measurements for simple partial covering fraction galaxy geometries (e.g. \citealt{Chisholm2018,Steidel2018,Saldana-Lopez2022}). 

\begin{figure}
\centering
\includegraphics[width=\columnwidth]{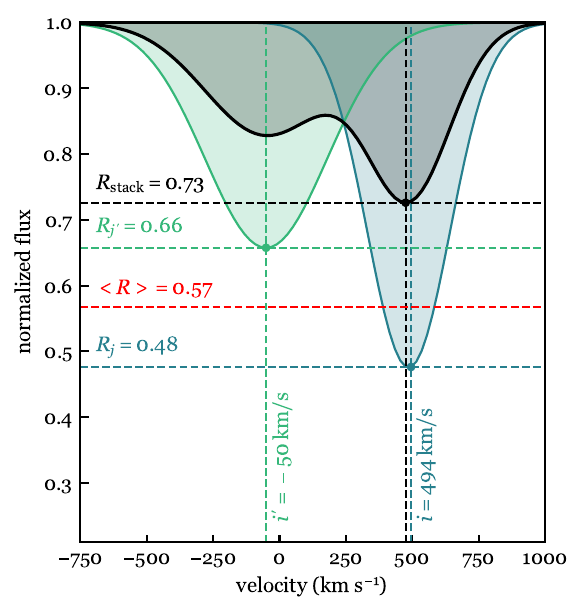}
\caption{A stacked spectrum of two sightlines, $j$ and $j'$, illustrating the effects on measuring residual flux. Due to the troughs of sightline $j$ and $j'$ not existing at the same wavelength bin, $i \neq i'$, the residual flux of the stacked spectrum does not equal the average of the constituent sightlines, $R_{\text{stack}} \neq <R>$.
\label{fig:R_diagram}}
\end{figure}

 Our analysis shows that even for the highest resolution, lowest noise spectral data, the process of stacking loses the residual flux information of the individual spectra. Interestingly, our worst performing stacks in relation to residual flux is the un-processed simulation spectra; up to 0.2 normalized units with a median of 0.06 in our presented cases. This is because residual flux measurements depend on the line shape. The residual flux is the flux measurement in a single, unique wavelength bin within an absorption line. When creating a stack, the wavelength bin position of the deepest point of the line in each spectrum is generally not the same. Figure~\ref{fig:R_diagram} shows how averaging two lines with different $v^{50}$ will always result in an underestimation of the $\avg{R}$. This is the same process of a single under-resolved line made up of multiple clouds within an individual sightline \citep[see Figure 17 of][for a further explanation]{Rivera-Thorsen2015}.

This effect can explain the apparent contradictory behavior seen among the different mock surveys in middle column of Figure~\ref{fig:all_stacks_2methods}. First, the diversity in the residual flux measurements in the simulation spectra represents the diversity in their line shapes. Thus the stacks tend to overstimate $\avg{R}$ by $\sim 0.06$. On the other hand, the LzLCS G140L, CLASSY, and the VANDELS surveys all do a better job of representing the underlying distributions in residual flux. This is due to a combination of reduced resolution and/or high S/N. Reducing the resolution by convolving the simulation spectra with a Gaussian kernel has the effect of homogenizing the line shapes that make up the stack. For the VANDELS stacks with poor resolution but decent S/N, all line shape diversity in constituent spectra is wiped out, causing $\Delta R$ to tend towards zero. Although $R_{\text{stack}}$ of the LzLCS G140L, CLASSY, and the VANDELS surveys may do a decent job of representing $\avg{R_{\text{obs}}}$, we remind the reader that this isolated effect is in \textit{addition} to the observational effects discussed in Section~\ref{sec:observational effects} and it may not be a good representation of the true average of $R_{\text{sim}}$.

Although it is clear why stacked spectra tend to overestimate residual flux compared to the average of the individual spectra, predicting by how much is unclear. The amount by which the residual flux measurement of the stack is off, $\Delta R$, is fully dependant on the unique line shapes of the constituent spectra. As a first step, \citet{Saldana-Lopez2023} reported residual flux correction factors when using stacks by doing similar tests using modelled Voigt profiles. For our simulated spectra that often have very complex and unique features, exploring the possibility of determining the line shape parameters that define the individual spectra from the stack is beyond the scope of this paper. This effect, in combination with how individual observational effects change residual flux measurements for a \textit{single} spectrum, makes reporting $\avg{R}$ from stacked spectra extremely unreliable and unpredictable without a deep understanding of the intrinsic line shapes of the individual spectra.

\subsection{Velocity}
The final column on the right of Figure~\ref{fig:all_stacks_2methods} shows the differences in the velocity measurements. The measurement $v^{50}_{\text{stack}}-\avg{v^{50}_{\text{obs}}}$ shows the difference between the average of the individual $v^{50}$ measurements of the constituent spectra and the single $v^{50}$ measurement on the stacked spectrum. Like equivalent width, the velocity differences are distributed evenly around zero, except for the LzLCS G140L and VANDELS survey with a median values of $-10$ and $-30$ \kms, respectively. The velocity residuals seem to be mostly affected by decreasing resolution and increasing noise. The stacks made from the LzLCS G140L mock data show the widest distribution with velocity residuals reaching $\sim 100$ \kms\ blueward and redward of zero. Although, it is important to note that these effects are small compared to the spectral resolution of each survey (see Table~\ref{tab:surveys}) and so we conclude that stacking effects on velocity measurements are minimal.

\section{Implications for Lyman continuum escape}\label{sec:wrv_discussion}

\begin{figure*}
\centering
\includegraphics[width=\textwidth]{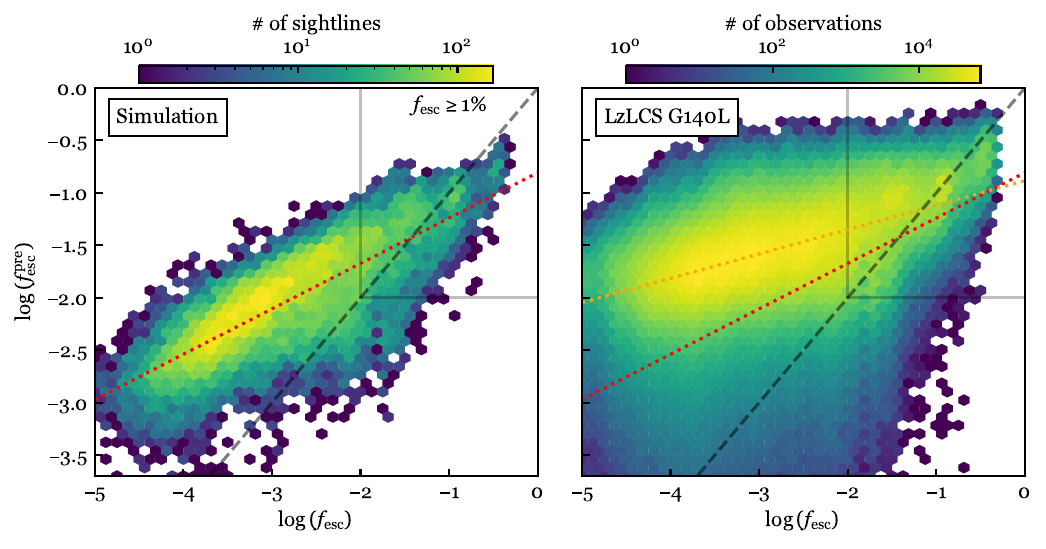}
\caption{Correlation between true LyC escape fraction and predicted escape fraction from dust corrected \SiII{} 1260  residual flux using Equation~\ref{eq:fpre}. {\bf Left:} The 22,500 true base sightlines are plotted for the simulation panel with a linear fit to the data shown by the dotted red line. {\bf Right:} The total 11,250,000 noise sample observations are plotted for the LzLCS G140L mock survey with the linear fit to the data shown by the dotted orange line. The upper right box shows where a LzLCS G140L observation of the \SiII{} 1260 residual flux  would correctly predict that a galaxy is a LyC emitter, with an escape fraction $\geq 1\%$.  In both panels, the grey dashed line indicates the 1:1 relation.}
\label{fig:fesc_R}
\end{figure*}

In Section \ref{sec:observational effects}, we quantified the impact of noise and spectral resolution on the measurements of equivalent width ($W$), residual flux ($R$), and velocity ($v^{50}$).  We created mock observations using the setups characteristic of COS galaxy surveys: the Low Redshift Lyman Continuum Survey (LzLCS), and the COS Legacy Archive Spectroscopic SurveY (CLASSY).  In some regards, the results of this exercise, highlighted in Figure \ref{fig:all_residuals} are not surprising.  While equivalent width is generally conserved in these COS observations, residual flux is significantly overestimated at the low-resolution of the LzLCS. 

Residual flux measurements have an important utility as a predictor of escaping LyC flux \citep{Heckman2011, Borthakur14, Jones13,Steidel2018,Gazagnes2020,Mauerhofer2021,Saldana-Lopez2022} with some of the first correlations between covering fraction and LyC escape in low redshift galaxies being made by \citet{Gazagnes2018} and \citet{Chisholm2018}. Subsequently, \citet{Saldana-Lopez2022} reported correlations between the LyC escape fraction and low ionization state (LIS) line measurements for the LzLCS sample observed with the G140L grating. 

For an assumed simplified picket-fence model with a uniform dust screen and in the limit of large optical depth in the covered regions
\begin{equation}\label{eq:fpre}
    f^{\text{pre}}_{\text{esc}} =  \frac{F^{\text{obs}}}{F^{\text{int}}} \times R = 10^{-0.4k_{912}E_{B-V}}R,
\end{equation}
where $F^{\text{obs}}$ is the observed flux in the line and $F^{\text{int}}$ is the intrinsic flux in the continuum around the line before dust attenuation. We extend the work of \citet[][see their Figure 22]{Mauerhofer2021} by including all simulation outputs, producing 22,500 measurements of LyC escape fraction, in order to explore the effects of LzLCS G140L resolution, binning and S/N. As described in \citet{Mauerhofer2021}, the LyC escape fraction is found by dividing the total number of ionizing photons that leave the virial radius by the total intrinsic ionizing photons. The total intrinsic ionizing photons are computed from mock spectra created for the wavelength range between 10\AA{} and 912\AA. Figure~\ref{fig:fesc_R} shows the correlation between predicted escape fraction via the measured residual flux and true escape fraction. 

First, we emphasize that even the high resolution simulation data tend to overpredict $f_{\text{esc}}$ at values $< 1\%$ and has significant scatter around $f_{\text{esc}} = f^{\text{pre}}_{\text{esc}}$ at values $\geq 1\%$. It has been shown that LIS lines like \SiII{} are not expected to perfectly trace \HI{} gas \citep{Henry2015,Gazagnes2018,Mauerhofer2021}, hence the spread and offset in the simulation data panel. Additionally, the assumed simplified geometry in equation~\ref{eq:fpre} is unlikely to be correct. As shown in \citet{Mauerhofer2021} and \citet{Gazagnes2023} the continuum of a line and the ionizing photons don't necessarily pass through the same media, making residual flux an unreliable indicator of escape fraction in certain cases. As we will discuss in section~\ref{sec:discussion_dust}, for our simulation, dust is not uniformly distributed within the sightline and is indeed clumpy, causing many complexities when inferring information from spatially integrated spectra. 

Second, the right panel of Figure~\ref{fig:fesc_R}  shows that $f^{\text{pre}}_{\text{esc}}$ made from mocking LzLCS G140L (orange dotted line) tends to severely over-predict the true LyC escape fraction (one-to-one line) and also exceeds the predicted value from the fiducial simulation data (red dotted line) for all but the highest $f_{\text{esc}}$ sightlines. Fits to our data show $\log{(f^{\text{pre}}_{\text{esc}})} = 0.43\log{(f_{\text{esc}})}-0.81$ for the fiducial simulation data and $\log{(f^{\text{pre}}_{\text{esc}})} = 0.23\log{(f_{\text{esc}})}-0.88$, a flattening of the slope by a factor of $\sim 2$, for the LzLCS G140L mock data. This is a direct consequence of the results presented in Section~\ref{sec:residual flux}. The left panel of Figure~\ref{fig:all_residuals_R} shows how the LzLCS G140L observations reveal measured residual flux values in excess of the true value for $\log{R_{\mathrm{sim}}} \lesssim 0.5$, translating to the over-prediction of LyC escape fraction. However, the average over-prediction is minimized as $R$, and consequently $f_{\text{esc}}$, increases. For sightlines with $f_{\text{esc}} \gg 1\%$, where determinations of LyC escape are most crucial to the questions of reionization, observational effects on $f^{\text{pre}}_{\text{esc}}$ like reduced resolution have minor effects on predictions. However, low S/N still introduces significant scatter. \citet{Saldana-Lopez2022} showed promising results by correcting LzLCS G140L $f^{\text{pre}}_{\text{esc}}$ measurements using the ratios between LIS and HI covering fractions. However, it is unclear whether this approach should be universally applicable as \citet{Mauerhofer2021} found that Ly$\beta$ residual flux measurements perform worse than LIS lines as a predictor of Lyman continuum escape. This is due to the fact Ly$\beta$ becomes optically thick at \HI{} column densities $\sim 3$ orders-of-magnitude lower than the LyC. This allows for high escape fraction sightlines where a large fraction of the area is optically thick to Ly$\beta$ but still optically thin to LyC.

\begin{figure*}
    \centering
    \includegraphics[width=\textwidth]{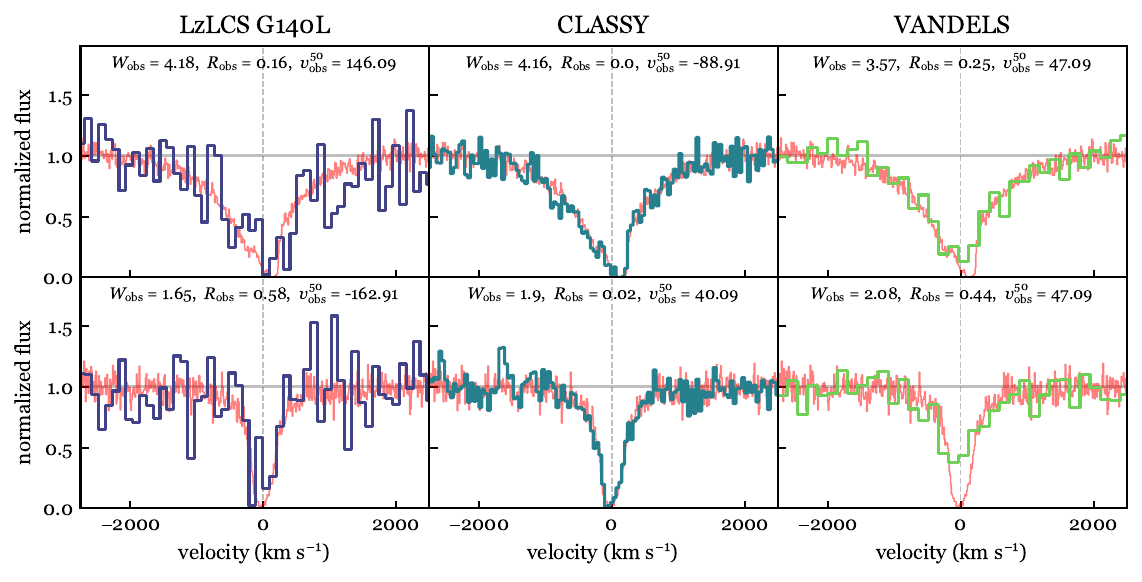}
    \caption{Same as Figure~\ref{fig:spec_examples} but for Ly$\beta$ $\lambda 1026$\AA{} with true line measurements: $W_{\text{sim}}=4.00,2.00$\AA, $R_{\text{sim}}=0.00,0.01$, $v^{50}_{\text{sim}}=-31.57,0.00$\kms.}
    \label{fig:spec_examples_lyb}
\end{figure*}

\begin{figure*}
\centering
\includegraphics[width=\textwidth]{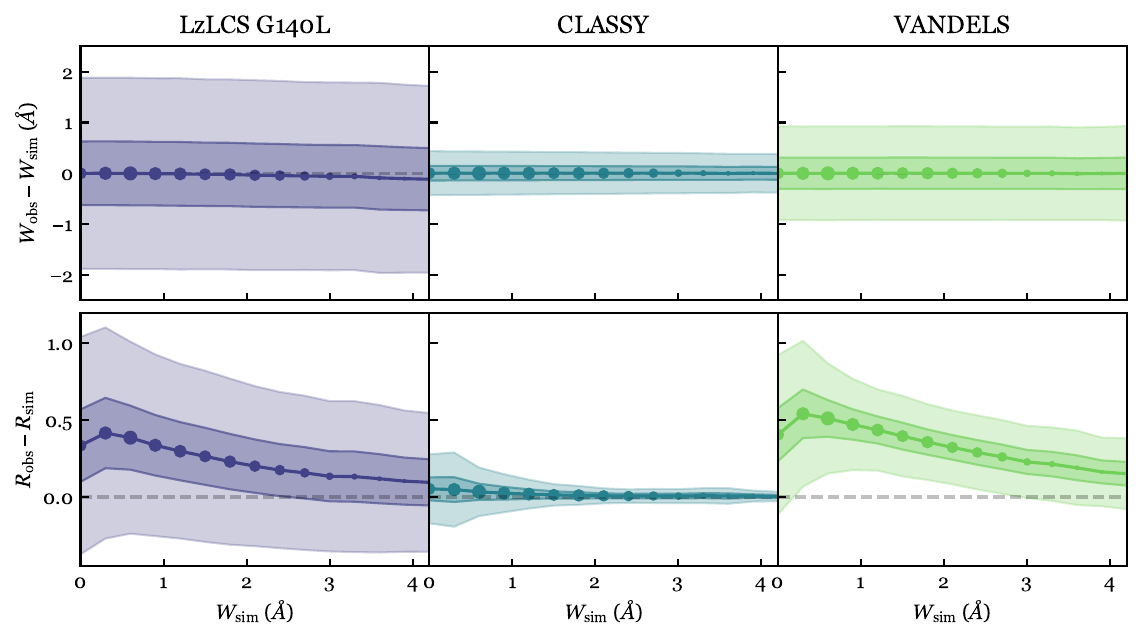}
\caption{Same as Figure~\ref{fig:all_residuals} but for Lyman$\beta$ $\lambda 1026$\AA.
\label{fig:all_residuals_lyb}}
\end{figure*}

More recently, \citet{Jaskot2024a,Jaskot2024b} used multivariate analysis to predict LyC emission, finding that line-of-sight probing diagnostics (absorption lines, Ly$\alpha$, dust) were more sensitive predictors than global properties (e.g. mass, the [OIII]/[OII] flux ratio). Although LIS (metal) line equivalent widths and residual fluxes from the LzLCS G140L observations were used in some predictors, they were found to be less significant than the same measurements for \HI{} when predicting LyC output \citep[as also discussed in][]{Saldana-Lopez2022}. Therefore, we also show our mock observational effects for the Ly$\beta$ $\lambda 1026$ \AA{} line in figures~\ref{fig:spec_examples_lyb} and \ref{fig:all_residuals_lyb}. The residual measurements for Ly$\beta$ show similar trends as previously discussed for the \SiII{} line with a larger spread in equivalent width residuals due to the larger integration width of -2750 \kms $\le v \le$ 2500 \kms\ needed for the broader line. The residual flux is again overestimated and we predict an average Ly$\beta$ residual flux overestimate of $\sim 0.16$ for the median measured equivalent width value of the LzLCS observations of $\sim 2.7$\AA{} \citep[see Table A.2 of][]{Saldana-Lopez2022}. While the \citet{Jaskot2024a,Jaskot2024b} works show a remarkable improvement in predictive ability of $f_{\text{esc}}$, it is still only accurate to $\sim 0.3$ dex when using the low resolution and S/N of the G140L observations. Ultimately, predictions from line measurements will be limited by the accuracy of such measurements in the LzLCS training set. This finding argues for observations to complete the high-resolution, high-S/N UV spectral library of LyC leakers and known non-leakers alike.

\section{\SiII{} Column Densities}\label{sec:column densities} 
The column density of gas, $N$, is a fundamental measure for the amount of gas along the line of sight in an absorption line system.  While nominally defined for point sources, column density is commonly also used for galaxies, particularly for the inference of mass-outflow rates that arise from star-formation driven winds \citep{Xu2022}.  
In detail, for a steady state spherical outflow, mass conservation requirements allow us to estimate the mass outflow rate of a galaxy,
\begin{equation} \label{eq:Mdot}
\dot{M} = \Omega v R_0 N_H m_p \mu,
\end{equation}
where $\Omega$ is the solid angle in steradians covered by outflowing gas seen in absorption, $v$ is the radial wind velocity, $R_0$ is the radial extent of the launch radius, $N_H$ is the total hydrogen column density, and $m_p\mu$ is the mean mass of particles. Using conventional analysis \citep[see][for a current alternative approach]{Carr2018,Carr2021}, ``down the barrel" galaxy spectra cannot uniquely constrain the three geometric outflow parameters ($\Omega,v,R_0$)\footnote{While velocity is not usually considered a ``geometric'' parameter, it should be noted that absorption line spectroscopy measures line-of-sight {\it projected} velocities, whereas an outflow may be directed towards a random, unknown orientation.}, but estimates put uncertainties for each within order unity. The total hydrogen column density, $N_H$, on the other hand, represents a large potential source of uncertainty. 

Using line-of-sight UV spectra, the current standard procedure for estimating $N_H$ involves deducing ionization fraction, elemental abundance and dust depletion factor of the ion through modeling \citep{Wang2021,Xu2022}, thereby introducing uncertainties at each step. In this section we focus on the first step of this calculation -- inference of the column density for the widely used \SiII{} ion. The results, however, are broadly applicable to many different ions.

\subsection{Simulation Column Densities}

\begin{figure*}
\centering
\includegraphics[width=0.95\textwidth]{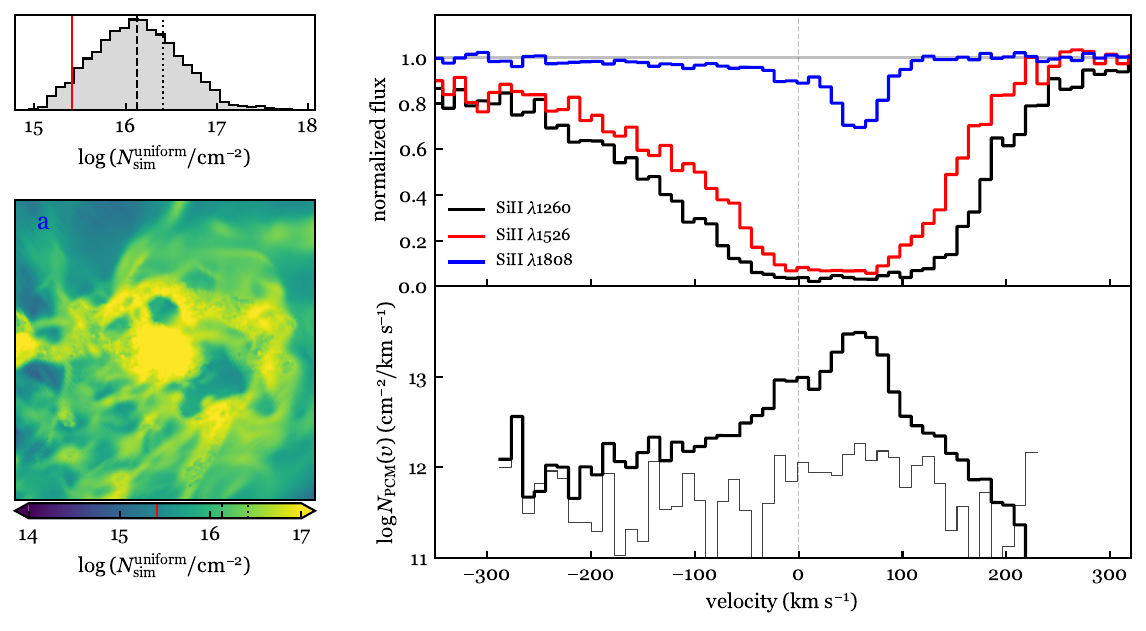}
\caption{\SiII{} column density diagnostics for a single representative sightline where the partial covering model produces an underestimation by $\sim 1$ order of magnitude. {\bf Left:} Column density map, 2kpc wide, with histogram of all pixel samples. The dashed and dotted vertical lines overlayed on the histogram and colorbar represent the median and mean column density values within the sampling aperture. The total inferred column density $N_{\mathrm{PCM}}$ is overlayed with a red solid vertical line. {\bf Right:} Observed spectra and inferred column density using the partial covering model. The thinner line shows the column density uncertainty in the PCM fit for each velocity bin.
\label{fig:N1_example}}
\end{figure*}

\begin{figure*}
\centering
\includegraphics[width=0.95\textwidth]{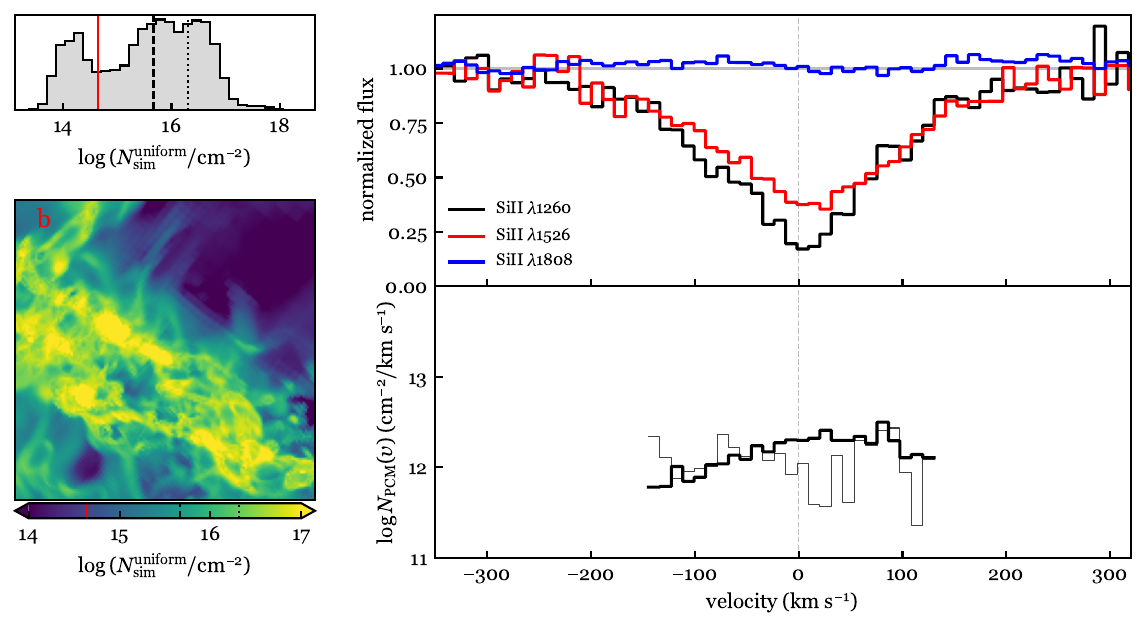}
\caption{Same as Figure~\ref{fig:N1_example} for a sightline where the partial covering model produces an underestimation by $\sim 1.5$ order of magnitude. 
\label{fig:N01_example}}
\end{figure*}

To produce the ``true'' column densities from the simulation for a given sightline we define a column with its base centered on the midplane of the galaxy and face perpendicular to the given sightline vector. The diameter of the column base is chosen at $\sim 2$kpc to encapsulate a majority of the stars. It is important to note that although this column size does not match the aperture size used to produce the spectra, this does not affect our findings. While the aperture size used to produce a spectrum is set by the observing instrument and the galaxy's apparent size on the sky, the correct column size used to calculate the column density is physically motivated by the geometry of the galaxy, i.e., where the stars are located. Including larger radii for the calculation of the simulation column would incorrectly add low density sight-lines that do not contribute to the absorption spectrum, since they are not back-lit by stars. Further justification for the chosen size is discussed in Appendix~\ref{sec:rprofile}. In addition to the aperture radius, we note that the length, $s$, of the column is extended from the mid-plane out to the virial radius. The column is divided into a $135 \times 135$ uniform grid of pixels that represent rays along which the column densities are computed,
\begin{equation} \label{eq:Nsim}
N_{\text{sim}} = \int n(s)\;ds,
\end{equation}
by integrating the \SiII{} particle densities, $n$, of the simulation cells. We therefore refer to this fiducial sampling method $N^{\text{uniform}}_{\text{sim}}$. Each column ``aperture" contains a distribution of \SiII{} column densities, as highlighted in Figures \ref{fig:N1_example} and \ref{fig:N01_example}. It is clear that attempting to report a single value of the column density allows multiple possibilities. For this paper we choose the mean of the aperture distribution for simulation column density values for each sightline, which should most closely resemble the total number of \SiII{} atoms divided by the aperture area\footnote{These are equal for an infinitely resolved spatial sampling. However, our uniform sampling introduces a slight bias towards diffuse, volume filling gas.}. Two example sightlines with their corresponding \SiII{} spectra and distributions in column densities are shown in Figures~\ref{fig:N1_example} and \ref{fig:N01_example}.

\subsection{Inferred Column Densities}
For calculating inferred column densities from the spectra our procedure follows the apparent optical depth method \citep{Savage1991} while assuming a partial covering model \citep[hereafter PCM;][]{Arav2005,Xu2018}. 
We solve the system of equations for the covering fraction and optical depth,
\begin{equation} \label{eq:I_k}
F_k(v) = 1 - C(v) + C(v) \, e^{-w_k \cdot \tau(v)},
\end{equation}
where $k$ stands for \SiII{} $\lambda \lambda 1260,1526,1808$, and the weight $w_k = f_k\lambda_k/f_{1260}\lambda_{1260}$ such that the solution for $\tau$ is for \SiII{} $\lambda 1260$. As will become apparent in Section~\ref{sec:line_sat}, it is necessary to include a line that is almost certainly optically thin. Therefore, we include the weak \SiII{} $\lambda 1808$ transition which provides the required dynamic range in $f\lambda$.

We solve equation~\ref{eq:I_k} using our three \SiII{} transitions in order to best fit $\tau(v)$ and $C(v)$ across the whole absorption line. The uncertainty on the flux levels are taken to be the measured standard deviation Monte Carlo noise at the continuum for each transition. Then the average ion column density in the aperture per velocity bin can be acquired via \citep{Savage1991,Arav2005,Rivera-Thorsen2015,Xu2022},
\begin{equation} \label{eq:Nion_v}
N_{\text{PCM}}(v) = \frac{m_ec}{\pi e^2 f\lambda}\avg{\tau(v)} = \frac{3.8\times 10^{14}}{f\lambda}\avg{\tau(v)},
\end{equation}
where $\avg{\tau(v)} = C(v)\;\tau(v)$ represents a spatially averaged optical depth assuming the partial covering model \citep{Edmonds2011}. The results are shown in the bottom right panels of Figures~\ref{fig:N1_example} and \ref{fig:N01_example}. The total estimate of the ion column density is then found by integrating across the whole line,
\begin{equation} \label{eq:Nion}
N_{\text{PCM}} = \int N_{\text{PCM}}(v)\;dv.
\end{equation}
We choose to integrate all velocity bins below the 80\% flux value of the absorption trough of the strongest \SiII{} $\lambda$1260 transition. This allows us to capture the majority of high column density velocity bins in the line core while avoiding excessively noisy portions toward the continuum. We take the native simulation spectra as our fiducial model when performing the PCM for inferred column densities. This allows us to isolate and evaluate the PCM's validity for estimated sightline column densities for galaxy observations. We discuss the outcome of mocking additional observational effects on PCM measurements later in Section~\ref{sec:line_sat}.

\subsection{Column Density Comparisons}\label{sec:column_density_comparisons}

\begin{figure*}
\centering
\includegraphics[width=\textwidth]{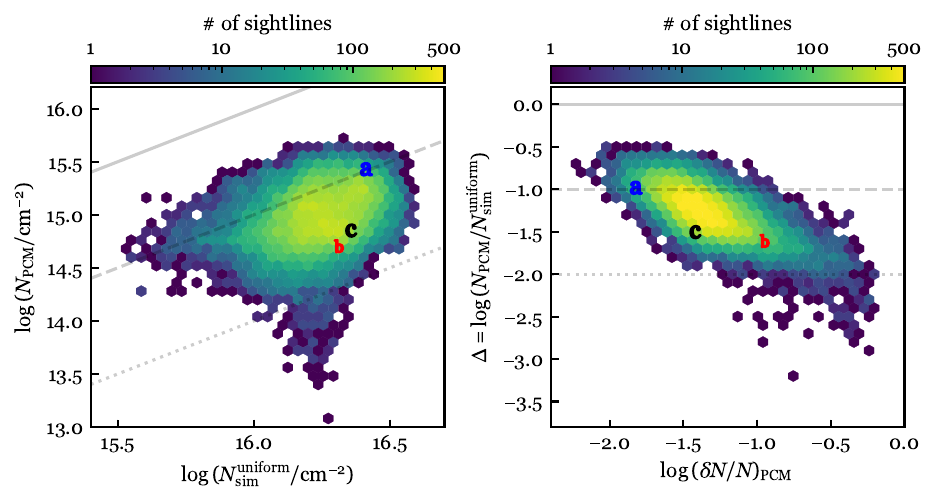}
\caption{A comparison of ``true'' and inferred SiII column densities for all 22,500 galaxy sightlines. The ``a'', ``b'' and ``c'' markers indicate the values corresponding to the sightlines shown in Figures~\ref{fig:N1_example}, \ref{fig:N01_example} and \ref{fig:N_methods} respectively. {\bf Left:} The simulation column densities are compared to those derived from the PCM using the \SiII{} $\lambda \lambda 1260,1526,1808$ transitions. On average the spectroscopically inferred \SiII{} column density is $\sim 1.5$dex lower than the \SiII{} column density attained by sampling the gas in the simulation cells over a uniform spatial grid. {\bf Right:} The column density underestimates are compared to the relative error on the PCM fits, showing a strong correlation between the two.
\label{fig:N_SiII_residuals}}
\end{figure*}

We compare simulation and inferred \SiII{} column densities with relative errors on the PCM fits in Figure~\ref{fig:N_SiII_residuals}. The $\log{(N^{\text{uniform}}_{\text{sim}}/\text{cm}^{-2})}$ distribution across all sightlines range from $15.5-16.6$ with the mode landing around $16.3$, while the $\log{(N_{\text{PCM}}/\text{cm}^{-2})}$ distribution has a larger spread from $13-15.5$ and the mode landing just below $15$. Meanwhile, a clear trend is seen in relative column density error, $\delta N$, with the bulk of PCM estimates showing $\log{(\delta N/N)_{\mathrm{PCM}}} \lesssim -1.0$.

The majority of inferred PCM column densities underestimate the simulation column density by $\sim 1$ to $1.5$ orders of magnitude with the median value being $\Delta = \log{(N_{\text{PCM}}/N^{\text{uniform}}_{\text{sim}})} = -1.25$. Figure~\ref{fig:N1_example} shows a representative sightline where the understimation is minimized around $\Delta \sim -1$. Figure~\ref{fig:N01_example} shows a representative sightline where the understimation is larger, around $\Delta \sim -1.5$.

\subsection{Reasons for column density discrepancies}\label{sec:column_discussion}
In Section \ref{sec:column_density_comparisons} we showed that \SiII{} column densities inferred from spectra using the PCM did not match the ``true'' measurements derived directly from the simulation. We now discuss the main reasons affecting observed column density estimates.

\subsubsection{Dust attenuation}\label{sec:discussion_dust}
\begin{figure*}
\centering
\includegraphics[width=\textwidth]{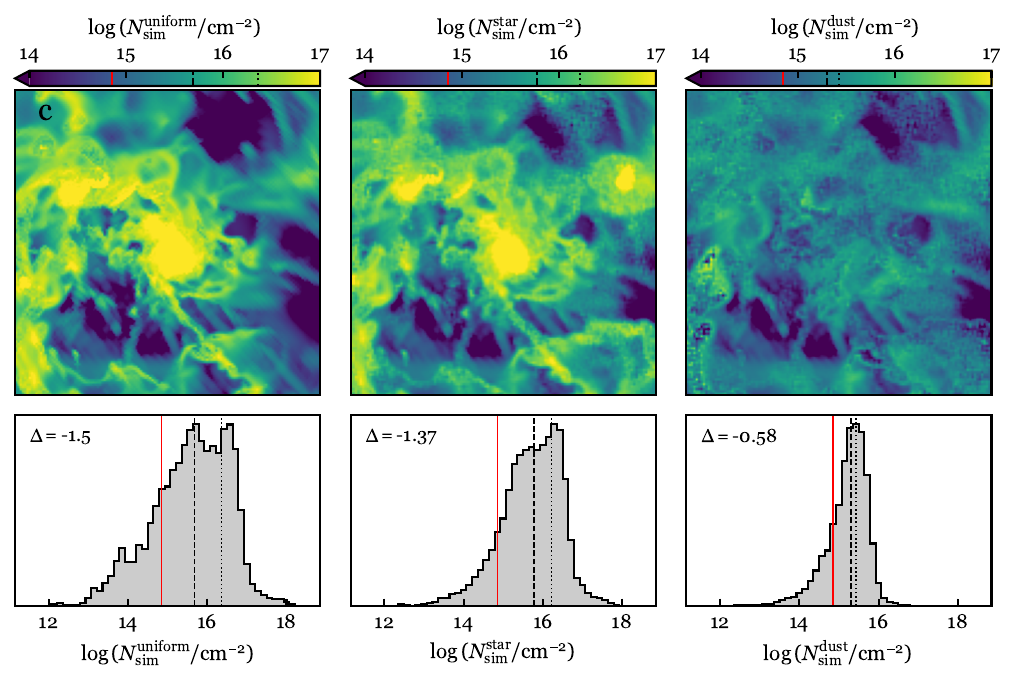}
\caption{\SiII{} column density diagnostics for a single example galaxy sightline using different sampling methods. From left to right, the three sampling methods are a uniform spatial sample, column densities in front of all stars, and column densities in front of all stars attenuated by dust. {\bf Top:} Column density maps for each sampling method. {\bf Bottom:} Histograms of ``true'' column densities for a single galaxy sightline for the three sampling methods. The vertical dotted and dashed lines show the mean and median values while the solid red line shows the PCM derived value. The uniform sampling and sampling in front of stars methods produce column densities $\sim 1.5$ order-of-magnitude greater than the observed column density. When dust attenuation is accounted for, the column density in front of stars is reduced by $\sim 1$dex and brings the observed column density within $\sim 0.5$dex.  
\label{fig:N_methods}}
\end{figure*}

In order to investigate the mismatch in reported column densities we employ different sampling methods from the simulation data. In contrast to our uniform spatial sampling method we choose to sample the \SiII{} column density directly in front of each star particle for a given sightline. As an absorption feature in a spectral line is created from backlit gas in the direction of the line of sight, the observed column density should be more representative of the gas backlit by stars, weighted by luminosity. Our star sampling method, $N_{\text{sim}}^{\text{star}}$, is compared to the uniform sampling method and shown in the middle column of Figure~\ref{fig:N_methods} for a chosen galaxy sightline containing $>10^6$ stars, and thus column density samples. Again, the mean column density value for this method is $> 1$ order-of-magnitude greater than the spectroscopically inferred value. 

Furthermore, high column sightlines tend to have high dust attenuation \citep{Guver2009} and although not implemented in this work, it has also been shown that dust depletion also depends on the average density along a line of sight, \citep{Jenkins09,Parvathi12,Choban22}. \citet{Mauerhofer2021} showed that the inclusion of dust can significantly increase the residual flux of LIS UV absorption lines (see their Figures 7, 9 and 10). Subsequently, \citet{Gazagnes2023} showed that the continuum and residual flux features can be produced from different spatial origins and that pixels with low dust attenuation contribute to $\sim90\%$ of the residual flux reflected in the spectrum. This allows the presence of dust to greatly increase the observed residual flux in a spectrum, leading to underestimates of the column density. 

In the simulation, regions of high density gas are cospatial with high amounts of dust (as modelled by the dust implementation; see Section~\ref{sec:data}) which should attenuate the light and cause a reduced measured flux in the spectrum. Therefore, we also calculate a dust column density, $N_{\text{dust}}$, in front of each star particle in order to estimate a simplified attenuated value for \SiII{} column density,
\begin{equation} \label{eq:Ndust}
N_{\text{sim}}^{\text{dust}} = N_{\text{sim}}^{\text{star}}*e^{-\tau_{\text{dust}}},
\end{equation}
where $\tau_{\text{dust}} = \sigma_{1260} N_{\text{dust}}$ and $\sigma_{1260}$ is an effective cross sectional area for dust grains around $\lambda$1260\AA{}. Our dust attenuated star sampling method, $N_{\text{sim}}^{\text{dust}}$, is seen in the right column of Figure~\ref{fig:N_methods}. It is clear when comparing the star sampling methods that the inclusion of dust attenuation should significantly suppress the highest column density samples, reducing the underestimation by $\sim 1$dex. Although this appears to improve the PCM underestimation problem to within $\Delta \approx -0.5$, $N_{\text{sim}}^{\text{dust}}$ does not represent the ``true'' amount of \SiII{} column density in front of the galaxy and is also an exaggeration of the effects of dust attenuation treating all dust to lie in a single plane in front of the \SiII{} gas. An in-depth treatment would require simulating another set of \SiII{} 1260, 1526, and 1808 absorption spectra with the dust optical depth channel deactivated and then used to fit PCM column density estimates. We will pursue this work for a future paper.

Current intrinsic dust correction procedures assume one of two models: a uniform dust screen or a clumpy model with dust only existing in the neutral gas clouds \citep{Scarlata2009,Borthakur14,Vasei2016,Reddy2016,Gazagnes2018}. In the former model, both radiation forming the continuum and absorption features in a spectrum pass through the same uniform dust screen and are attenuated by the same amount, having no effect on the normalized absorption profiles of lines. However for the latter model, only a fraction $C$ of radiation passing through the neutral gas clouds becomes attenuated by dust while the rest leaves freely. In this case, dust can greatly affect the normalized absorption profile of the lines with the residual flux depending on an additional factor $10^{0.4k_{\lambda}E_{\text{B-V}}}$.

Distinguishing between the two models from 1D spectra alone can be challenging. The effect of dust on absorption line residual flux values for the clumpy model should manifest itself across multiple transitions of the same line due to the wavelength dependence of the attenuation law. Unfortunately, resolution and S/N requirements for confirming this effect may be prohibitive. \citet{Gazagnes2018} estimated resolutions of $R = 15,000$ and S/N = 30 are necessary to distinguish between dust models using Lyman series lines. For this reason, it is common to assume a uniform dust screen model due to its simplicity over a clumpy model, especially when estimating LyC escape fractions. We show here that assuming a uniform dust screen model for systems that are actually clumpy can lead to large underestimates of the column density via the PCM.

\begin{figure}
\centering
\includegraphics[width=\columnwidth]{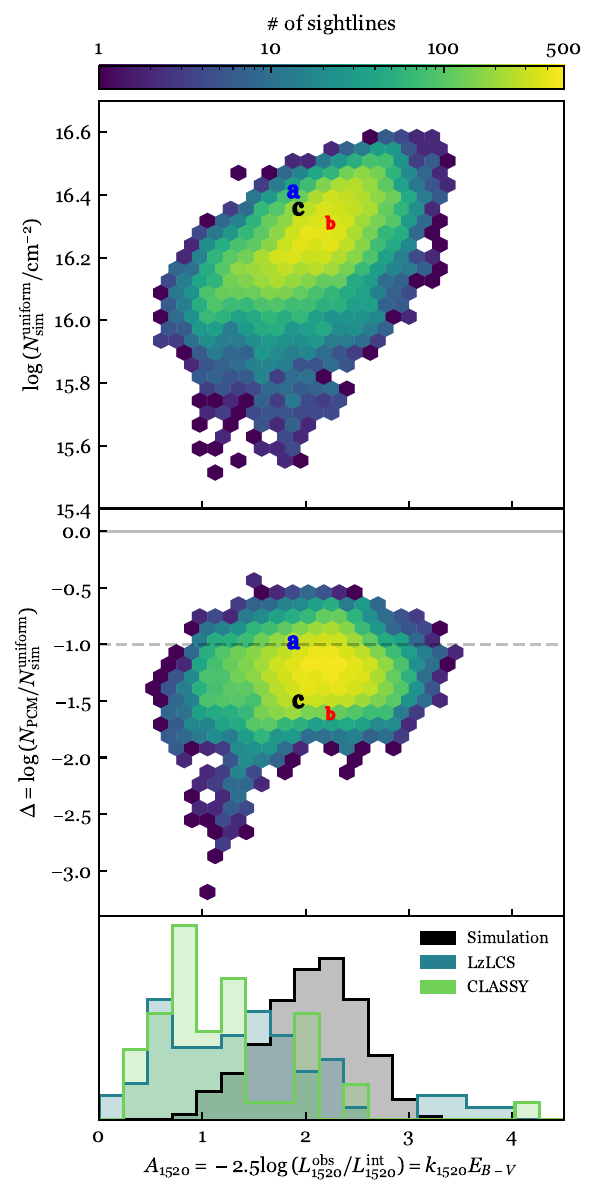}
\caption{{\bf Top:} True \SiII{} column density as a function of integrated sightline UV attenuation at $1520$\AA{} for all 22,500 galaxy sightlines. {\bf Middle:} \SiII{} column density underestimates as a function integrated sightline UV attenuation. {\bf Bottom:} UV attenuations for the simulation sightlines, LzLCS, and CLASSY galaxies. A large fraction of the LzLCS and CLASSY samples overlap with the simulated galaxy sightlines.
\label{fig:extinction}}
\end{figure}

In Figure~\ref{fig:extinction} we plot the \SiII{} column densities as a function of integrated sightline UV attenuation at $1520$\AA, $A_{1520} = -2.5\log{(L^{\text{obs}}_{1520}/L^{\text{int}}_{1520})}= k_{1520}E_{\text{B}-\text{V}}$, where $L^{\text{int}}$ is the intrinsic galaxy luminosity, and $L^{\text{obs}}$ is the galaxy luminosity as observed within the aperture with dust included. In the top panel we can see that $N^{\text{uniform}}_{\text{sim}}$ and $A_{1520}$ increase in tandem as expected by the dust implementation explained in Section~\ref{sec:data} and consistent with a clumpy dust model. In the middle panel we plot the \SiII{} column density underestimates as a function of the UV dust attenuation. This shows how the integrated UV attenuation does not map to a single value of $\Delta = \log{(N_{\text{PCM}}/N^{\text{uniform}}_{\text{sim}})}$. For example, points "a" and "c" share similar $N^{\text{uniform}}_{\text{sim}}$ and $A_{1520}$ values but differ by $\approx 0.5$ in $\Delta$. This highlights the complex effect dust can possibly have on the inferred column density using the PCM because the observed residual flux of absorption lines for a clumpy dust model is a function of optical depth, covering fraction \textit{and} the amount of dust attenuation \citep{Gazagnes2018,Chisholm2018}. Consequently, if a uniform dust screen model is incorrectly assumed for a clumpy ISM and outflow like our simulated galaxy, PCM inferred column densities can be severely underestimated. Therefore, it is apparent that a full understanding of the spatial distribution of the gas and dust within a galaxy is necessary to accurately estimate column densities via the PCM. 

\SiII{} column densities have been estimated for CLASSY \citep{Xu2022}, and are possible to estimate for the galaxies of LzLCS that have been observed with sufficient signal-to-noise in higher resolution than the G140L grating provides. The bottom histogram of Figure~\ref{fig:extinction} compares the range of UV attenuations of the simulated galaxy sightlines with the LzLCS and CLASSY sample galaxies. Although the UV attenuations of the simulated galaxy sightlines tend to be slightly higher than the survey galaxies, there is still considerable overlap. Therefore, our findings may be applicable to these survey galaxies. However, the spatial location of the dust may determine the severity of the effect on inferred column density estimates, e.g. dust primarily located in the ISM will have limited effect on column density estimates in the wind. High resolution integral field unit (IFU) spectroscopic data cubes of the survey galaxies in multiple wavelength bands would help confirm or rule out these findings by allowing a more robust understanding of the dust distribution within a spatially resolved sightline. 
    
\subsubsection{Optical depth averaging}\label{sec:tau_avg}
The spectra produced in this investigation result from a combination of pixels or spatial simulation bins (a spatially extended object), as illustrated in Figures \ref{fig:N1_example}, \ref{fig:N01_example} and \ref{fig:N_methods}. Additionally, since the \SiII{} column is not uniformly backlit by stars, the resulting spectrum is weighted by the brightness of the background source. In detail, the normalized flux of an absorption line within an aperture of area $A=x*y$ is
\begin{equation} \label{eq:flux}
F(v) = \frac{\int S(x,y)e^{-\tau(v;x,y)}\,dx dy}{\int S(x,y)\,dx dy} = \avg{e^{-\tau(v)}},
\label{Eq:weighted_flux}
\end{equation}
where $S$ is the background source brightness\footnote{This mathematical representation of an absorption line system is fully consistent with a partial covering model, but does not distinguish spaxels with $\tau(x, y, v) = 0$ or $\infty$ from those with finite optical depth. Hence, the covering fraction, $C_f(v)$ is unity in this case.}. This equation can be thought of as a data cube, analogous to an integral field unit spectrum, where each spaxel can be analyzed to derive $\tau(v)$ in each velocity bin.  It follows, then, that we can measure $\avg{\tau(v)}$ by averaging over the spaxels.  
However, when combining spaxels into a single spectrum, the resulting spectrum {\it does not} encode the average optical depth $\avg{\tau(v)}$. Rather, following Equation \ref{Eq:weighted_flux}  the spectrum yields an \textit{apparent} optical depth, $\tau_a \equiv \ln{(1/\avg{e^{-\tau}})}$ \citep{Savage1991}. According to Jensen's inequality, $\avg{e^{-\tau}} \geq e^{-\avg{\tau}}$; therefore, assuming that a spectrum has yielded the right hand side of the inequality, when in fact it is the left hand side will always result in an underestimate of optical depth and column density. 

The full distribution of the optical depth information is consequently lost when a spectrum is created and there is no clear path to recover $\avg{\tau(v)}$ from the observationally derived $\tau_a(v)$. \citet{delaCruz2021} emphasised the consequence of this mathematical effect on mock spectra created from a high resolution cloud-wind simulation. Adopting a constant background source and neglecting the effect of dust, they highlighted the isolated effect of the optical depth averaging and showed that observations can under-predict the true value for optically thick ions.

\citet{Huberty2024} also explored this effect by comparing \SiII{} column densities estimated via the PCM with spectra created using a Semi-Analytical Line Transfer \citep[SALT;][]{Scarlata2015,Carr2018,Carr2021,Carr2023} model and found that the PCM underestimated the true value. Again these tests were done on gas that was uniformly backlit, spherically symmetric (although biconical geometries were allowed) and did not explore $S$ as a function of sightline. In this paper, the background source brightness is instead a function of location within the aperture, $S(x,y)$, due to the luminosity of individual star particles and we confirm the expected underestimation of \SiII{} columns via the PCM persists (see the middle column of Figure~\ref{fig:N_methods}). In addition, our simulations are not spherically symmetric and show outflows that are highly variable within a given sightline.

\begin{figure*}
\centering
\includegraphics[width=\textwidth]{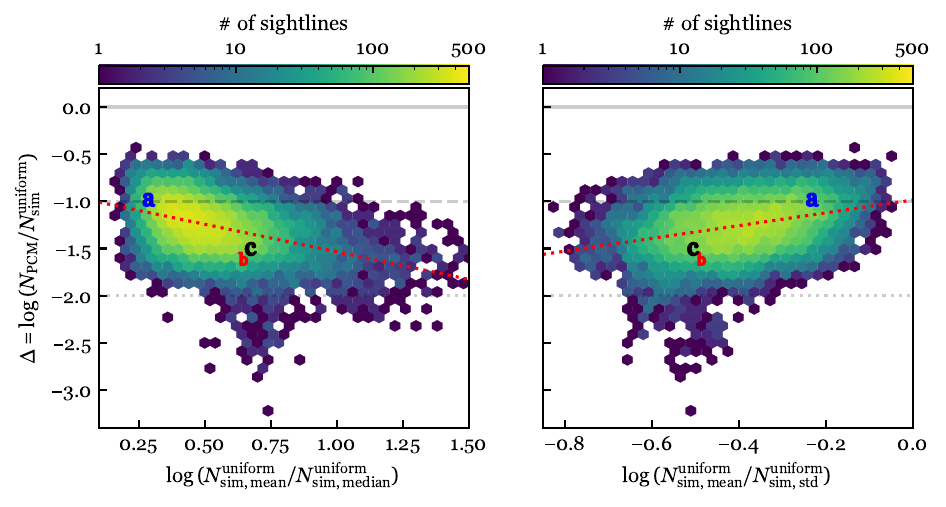}
\caption{Relationships between column density underestimates and spatial column density distribution statistics. The ``a'', ``b'' and ``c'' markers indicate the values corresponding to the sightlines shown in Figures~\ref{fig:N1_example}, \ref{fig:N01_example} and \ref{fig:N_methods} respectively. {\bf Left:} The ratio of the mean to median column density within the simulation aperture. The ratio is related to the skewness of the column distribution. $\Delta$ decreases as the column density distribution skews more negative in logarithm space. Larger ratios also encode multimodal distributions with more holes as shown in Figure~\ref{fig:N01_example}. {\bf Right:} The ratio of the mean to standard deviation of the column density within the simulation aperture. The ratio is the coefficient of variation and shows how as the variation of column densities increases (to the left) the PCM has a harder time recovering the mean column density.
\label{fig:Delta_stats}}
\end{figure*}

As a proxy for optical depth we explore the \SiII{} column density spatial distributions within each aperture sightline. In Figure~\ref{fig:Delta_stats} we attempt to quantify the distributions by comparing the mean column density to the median, $\log{(N_{\text{mean}}/N_{\text{median}})}$, and standard deviation, $\log{(N_{\text{mean}}/N_{\text{std}})}$, within the aperture and how it affects the error in the PCM fits. The ratio of the mean to median is related to the skew of the column density distribution and can reveal possible multimodal characteristics; in other words, sightlines with more prominent holes. The ratio of the mean to standard deviation shows the variation in the distribution. Both of these statistics show correlation with the PCM column density understimates, with $\Delta$ becoming worse by upto $1$dex as variation in the column distribution increases. Fits to our data show $\Delta \propto -0.58\log{(N_{\text{mean}}/N_{\text{median}})}$ and $\Delta \propto 0.68\log{(N_{\text{mean}}/N_{\text{std}})}$, with the y-offsets most likely being determined by dust effects. This behavior suggests that the PCM has a difficult time recovering the true column density for sightlines with high spatial variation in column density. One of the main assumptions of the PCM is that the geometry can be explained by the source either being fully covered (optically thick) or not (optically thin), at a given projected velocity and that the covered regions are saturated with the same optical depth \citep{Arav2005}. This description may not be a very good representation of realistic sightlines of galaxies with active outflows.
 
\subsubsection{Line Saturation and Finite Spectral Resolution}\label{sec:line_sat}
It is generally understood that line saturation governs the ability to constrain column density from absorption lines. Since lines with non-zero residual flux may still be saturated if they only partially cover the background source, the PCM uses multiple transitions in the same ion to solve for $C_f(v)$ and $\tau(v)$.  For optically thin lines, the strength of the absorption is proportional to $f\lambda$, whereas saturated lines will have absorption profiles that appear to have the same strength regardless of $f\lambda$.  Critically, however, LIS lines are often {\it neither} optically thin nor saturated; intermediate optical depths around $\tau(v) \sim 1$ are common both in the simulation data presented here, and in real observations \citep{Xu2022}.  In these cases, observations of the \SiII{} 1260 and 1526 lines, appearing like those in Figures~\ref{fig:N1_example} suggest that the PCM is applicable.  However, similar to the spatial optical depth averaging,  when integrating  over the line profile, optical depths (and column densities) do not add linearly. As pointed out in \cite{Huberty2024}, when this becomes the case, the results depend on the spectral resolution.  Only for truly optically thin absorption, when $\tau << 1$ and  $e^{-\tau} \approx 1 - \tau$ the column density does not depend on resolution.   In this case, the column density is proportional to equivalent width, which is conserved with decreased resolution.  It is for this reason that we chose to add the \SiII{} 1808 line, which is 45 times weaker than \SiII{} 1526.

We tested this effect using simple Voigt profiles \citep[approximated according to][]{Smith2015}  to model the lines. 
We fit the PCM to determine the \SiII{} column density from the $\lambda$1260, 1526, and 1808 transitions with Doppler parameters of $b = 25$\kms (increasing the Doppler parameter increases the column density at which the transitions saturate). For the theoretical profiles, the PCM recovers the true column density for $\log{(N_{\mathrm{true}}/\text{cm}^{-2})} \lesssim 14.5$. For $\log{(N_{\mathrm{true}}/\text{cm}^{-2})} \gtrsim 14.5$, two of the three transitions used to fit the PCM become saturated in the core of the absorption trough, thereby reducing the constraints on the optical depth. This causes the PCM to underestimate the true column density. At a column density of $\log{(N/\text{cm}^{-2})} = 16$ the PCM underestimates the theoretical value by $\sim 0.08$dex, a negligble amount compared to our findings with the simulated data. Unlike single Voigt profiles, the absorption lines produced from the simulation are complex combinations of multiple outflowing and inflowing components that allow for larger velocity ranges showing saturation which may lead to even larger underestimates.

\begin{figure}
\centering
\includegraphics[width=\columnwidth]{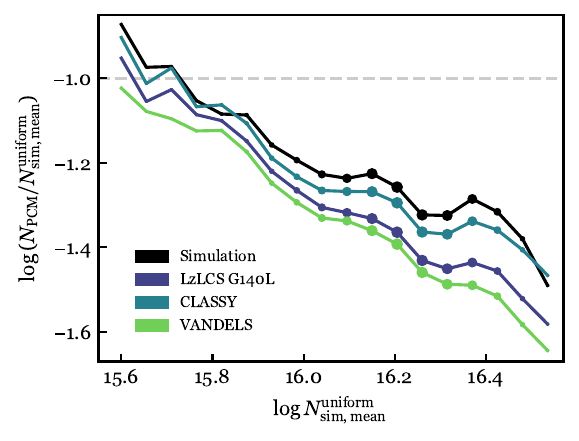}
\caption{A comparison of ``true'' and inferred SiII column densities for the different spectral resolutions and binnings of each mock survey. Sightline data is binned by $\log{N_{\text{sim}}}$ and the median value per bin is shown. Marker sizes are proportional to the sightline counts per bin.
\label{fig:res_tests}}
\end{figure}

\citet{Huberty2024} explored this effect on outflow spectra produced with SALT. They found that reduced resolution alone can explain PCM \SiII{} column density underestimates by up to $\sim 1$dex for $\log{(N/\text{cm}^{-2})} \gtrsim 16$ at the resolution of our simulated spectra of $\sim 10$\kms. Although our findings are generally consistent for our fiducial resolution (see Figure~\ref{fig:N_SiII_residuals}), we further test the effects of reduced resolution on the estimated PCM column density. We downgrade the resolution and binning of our spectra to match our mock surveys and plot the resulting PCM estimates in Figure~\ref{fig:res_tests}. We choose not to add the mock noise for each survey as it complicates the interpretation and only introduces a larger spread in the distribution of measurements. While this figure shows the overall trend of more drastic PCM failure at high column densities, there is minimal difference from the change in resolution when directly comparing to \citet{Huberty2024}. At $\log{(N/\text{cm}^{-2})} = 16$, \citet{Huberty2024} predict a increase in magnitude of $\Delta$ by $\sim 0.4$ dex when downgrading the spectral resolution from 10\kms to  60\kms. For our simulated data we only see an increase in magnitude of $0.1$ when reducing the resolution from our fiducial simulated spectrum of 10\kms all the way down to the VANDELS' 461\kms. 

This difference highlights the role that individual line shapes play in the robustness of the PCM estimates. We note that the downgrading of spectral resolution has a greater ability to ``hide'' saturation for lines that have narrower cores and less pronounced wings, as in the case for isolated outflow spectral lines. In the case of the modeled absorption lines in \citet{Huberty2024} \citep[which can be seen in][]{Carr2023}, steep flux gradients are found on the redward side of the absorption trough; essentially holding a large proportion of saturation in a single or a few pixels at $\sim 0$\kms. This is due to modeling an isolated homologous outflow in front of a point source. With the ISM component removed, a steep gradient in the absorption flux for velocities $>$0 \kms{} occurs. Additionally, SALT does not include line broadening represented by the Doppler parameter (i.e. thermal and turbulent broadening) which would further reduce resolution effects. Therefore, the results of \citet{Huberty2024} highlight the isolated effect of resolution on pure outflow PCM measurements and help inform a worst case scenario for column density estimates. On the other hand, a majority of our simulated absorption lines are quite broad with smooth transitions between the core and wings on both the blue and red sides due to complex outward, inward and turbulent flows, as well as the inclusion of the ISM. For this reason we expect resolution effects to be less pronounced than in \citet{Huberty2024}. However, the severity of this effect may be specific to the galaxy and outflow observed, with the findings of \citet{Huberty2024} and this paper representing two possible outcomes. Consequently, understanding the true underlying absorption line shapes is vital to estimating the effects of resolution on PCM column density estimates.

\subsection{PCM on Stacked Spectra}\label{sec:stacking_discussion}
\label{sec:discussion_stacking}

\begin{figure}
\centering
\includegraphics[width=\columnwidth]{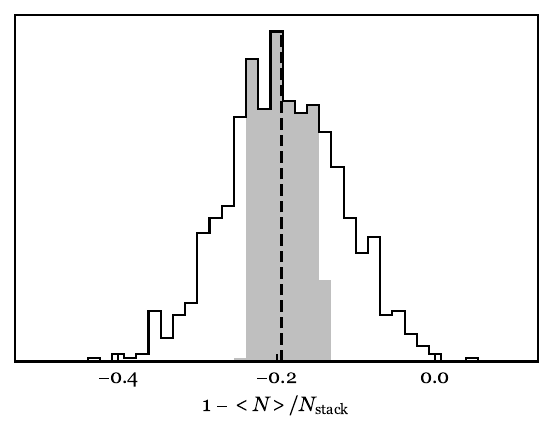}
\caption{Outcome of 1000 measurements comparing PCM column densities from stacked spectra $N_{\text{stack}}$ to the average PCM column densities of the individual sightlines $\avg{N}$. The vertical dashed line shows the median outcome where $N_{\text{stack}}$ underestimates $\avg{N}$ by 20\%.
\label{fig:N_Nstack}}
\end{figure}

In light of stacked spectra overestimating the average flux shown in Section~\ref{sec:stacking effects}, we additionally test the effects on column density estimates using the PCM. We make column density estimates from 1000 random sets of 30 stacked \SiII{} $\lambda 1260$, $\lambda 1526$ and $\lambda 1808$ spectra, and compare the stack measurements to the average values obtained from the constituent spectra. The results are shown in Figure~\ref{fig:N_Nstack}. Using stacked spectra results in an underestimate  of the average of column density values by around $20\%$.  This bias is \textit{in addition} to the previously discussed $\sim$1.5 dex underestimate that is inherent in the PCM measurements from the individual spectra. This result dissuades the use of the PCM to estimate column densities on stacked spectra. However, with reliable knowledge that the individual spectra contain optically thin lines, estimates of column densities using equivalent width analysis is feasible as was done for sulfur lines in \citet{Xu2022}. As we showed in Section~\ref{sec:stacking effects}, equivalent width is conserved with stacking, provided that the individual spectra are continuum-normalized.

\section{Summary}\label{sec:summary}
Rest-frame UV spectroscopy, in the form of line-of-sight absorption and emission lines, provides a means of unveiling crucial information about a galaxy's ISM and CGM. However, despite being a main tool, limitations inherent in observing these line features like resolution and S/N can make interpreting measurements difficult. By using a simulated galaxy as a controlled laboratory, we created mock \SiII{} spectra via radiative transfer to test the observational effects on common UV absorption line measurements and the fidelity of spectral stacking and the partial covering model.

By directly comparing our fiducial simulated \SiII{} spectra against the LzLCS G140L, CLASSY and VANDELS mock survey spectra, we were able to better understand the effects of reduced resolution and degraded S/N on common absorption line measurements. We find that residual flux measurements of individual line observations are the most significantly affected by reduced resolution and S/N. As expected, a reduction in resolution redistributes power in the core of a line into the wings, causing the apparent residual flux to be greater. For the LzLCS G140L mock spectra, residual flux measurements are on average $\sim 300\%$ greater than the true value for the deepest lines where $R = 0.1$ (see Figures~\ref{fig:all_residuals} and \ref{fig:all_residuals_R}). Even for the highest resolution and S/N observations of CLASSY, residual flux is on average overestimated by $\sim 0.1$ normalized flux units for absorption lines with equivalent widths $<1$\AA. Equivalent width and velocity measurements are on average conserved, while the scatter increases towards lower S/N.

We then explored the commonly employed technique of spectral stacking by directly comparing measured line properties on stacked spectra against the average values of the constituent set. We find that residual flux measurements of a stacked spectrum are the most significantly affected by the process of stacking. We find that a stacked spectrum does not retain information on the residual flux of the individual constituent spectra which makes decoding the underlying distribution of line residual fluxes challenging and unpredictable. On average, the residual flux of a stacked spectrum overestimates the average of the individual residual flux measurements. We find the largest difference in the highest resolution fiducial simulation spectral stacks, with $R_{\text{stack}}$ on average being 0.06 normalized units larger than the average $<R>$ of the set (see Figure~\ref{fig:all_stacks_2methods}). The CLASSY mock stacks showed a median value of $R_{\text{stack}} - <R_{\text{obs}}> = 0.02$. However, as the spectral resolution of observations continues to be degraded, variations in the line shapes are minimized, and the effects of stacking become less unpredictable. Therefore, we recommend not using stacks to estimate average residual flux and column density measurements using the partial covering method. However, provided that spectra are normalized prior to stacking, the equivalent width is preserved. Stacks of un-normalized spectra yield equivalent width measurements that are biased towards the brighter objects. As such, the equivalent width from un-normalized stacks represents a flux-weighted mean value which is usually unknown when stacking methods are employed. Therefore, we advocate for the use of normalized stacks. Lastly, we find that differences in velocity measurements $v^{50}_{\text{stack}} - <v^{50}_{\text{obs}}>$ fall well within the velocity resolution and binning of the respective mock survey.

We also explored whether the simulated spectra could recover the known column density of \SiII{} in the simulation. We find that the partial covering model applied to our fiducial spectra of \SiII{} 1260,1526 and 1808 absorption lines, systematically underestimates the ``true'' \SiII{} column density value $\Delta = \log{(N_{\text{PCM}}/N^{\text{uniform}}_{\text{sim}})} = -1.25$ on average (see Figure~\ref{fig:N_SiII_residuals}). We explored three possible culprits for this discrepancy: (1) dust attenuation, (2) optical depth spatial averaging, (3) hidden saturation by reduced resolution. We conclude that dust attenuation has the highest potential for causing underestimates of \SiII{} column densities via the partial covering model for our simulation. By weighting the column density in front of each star particle by the amount of dust we find that dust can be responsible for at least $\sim 1$ dex of our PCM under-estimates (see Figure~\ref{fig:N_methods}). In other words, the lines-of-sight with the highest column densities of metal ions (like \SiII{}) will also generally have the highest dust opacity. The material in these sight-lines will essentially be invisible in the UV and will not contribute to the observed absorption-line profiles. Most importantly, the effect of dust on the residual flux of the absorption lines used for the PCM estimates is highly dependent on the dust geometry of the ISM and outflow. If a uniform dust screen is assumed when in fact a clumpy model is more accurate, this effect will be missed and column estimates may be greatly deficient. 

Next, with the understanding that optical depths will always be underestimated within a spectrum due to spatial averaging, we investigated the spatial distributions of \SiII{} column densities. We find that the PCM under-estimates are correlated with the skew and standard deviation of the spatial column density distributions; sightlines with larger skew and variance performing worst (see Figure~\ref{fig:Delta_stats}). We find $\Delta \propto -0.58\log{(N_{\text{mean}}/N_{\text{median}})}$ and $\Delta \propto 0.68\log{(N_{\text{mean}}/N_{\text{std}})}$. Lastly, we recalculated PCM column density estimates after downgrading the spectral resolution from our fiducial simulation value of 10 \kms\ to match those of the CLASSY (65 \kms), LzLCS G140L (300 \kms) and VANDELS (461 \kms) surveys. We find our PCM under-estimates at $\log{(N/\text{cm}^{-2})} = 16$ worsen by only 0.1 dex in total, signifying that our \SiII{} absorption lines are sufficiently resolved for the use of the partial covering model.

This study provides key insights into the use and limitations of UV spectroscopy of star forming galaxies.
We have identified and explored three main phenomena that have the potential to cause significant column density underestimates measured using the partial covering model. We highlighted attenuation via spatially distributed dust as a key component in underestimates of \SiII{} column densities. However, to truly disentangle this effect from optical depth averaging the isolation of the two is important. In a future paper we will generate spectra with dust turned off to better quantify its sole effect. While weighting simulated optical depths by the amount of dust column in front of each star provides a better match to the PCM-derived value, this information is not available in single aperture spectra. To reliably estimate the true amount of gas column along a galaxy sightline, a detailed knowledge of the spatial distribution of stars, gas and dust is required. This investigation strongly motivates the inclusion of a UV integral-field-unit spectrograph in next generation space-based observatories.

\begin{acknowledgements}
AH and RMJ acknowledge support from HST GO 15626, and the STScI director's discretionary research fund.  We are grateful to John Chisholm, Harley Katz, and Bethan James for fruitful discussions. 
\end{acknowledgements}



\appendix

\section{Aperture Effects on Inferred Column Density}\label{sec:rprofile}

\begin{figure}
\centering
\includegraphics[width=\textwidth]{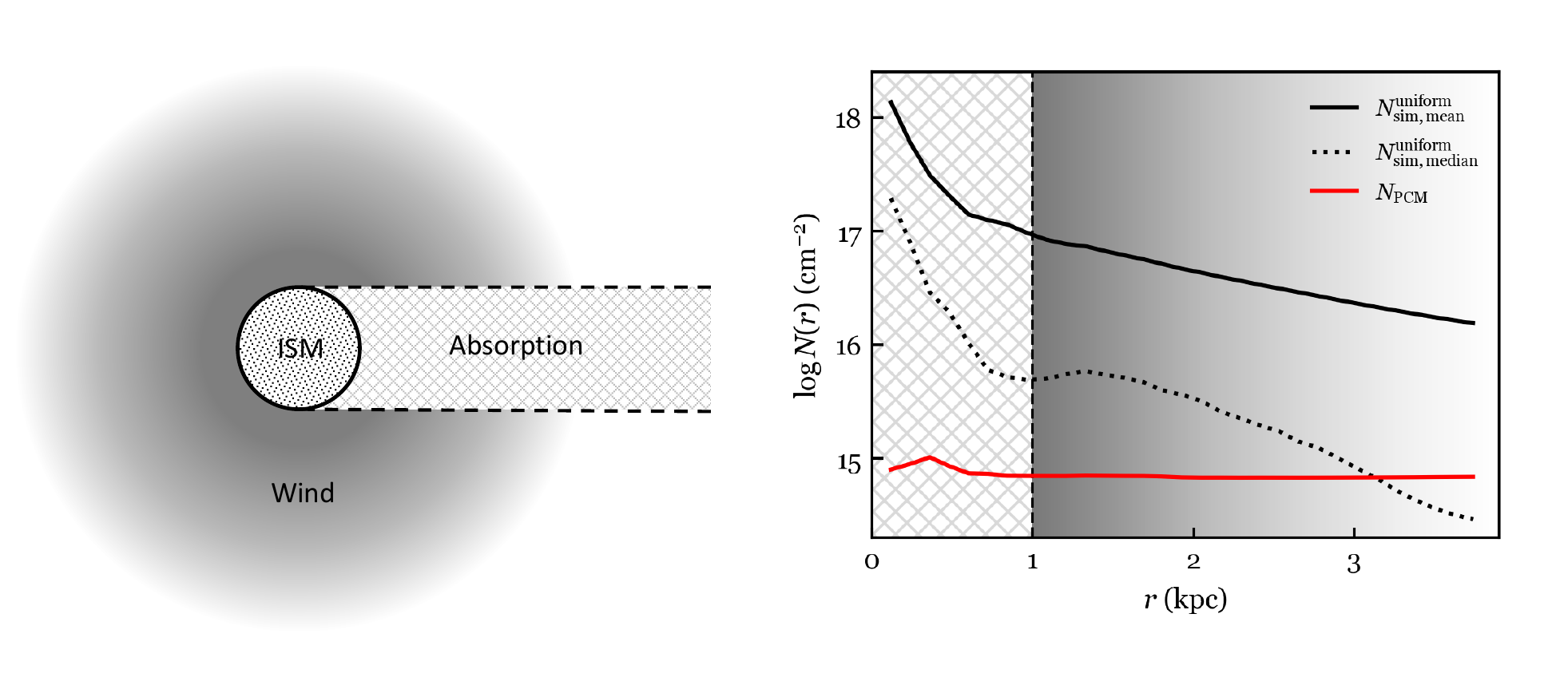}
\caption{\textbf{Left:} Galaxy diagram showing a central dense ISM surrounded by a more diffuse extended wind. The region between the stellar continuum in the ISM and the observer is labelled as the absorption region. The absorption feature in the spectrum used for the PCM originates from this region. \textbf{Right:} Column density measurements as a function of aperture radius for a single representative sightline. Both $N_{\text{sim,mean}}(r)$ and $N_{\text{sim,median}}(r)$ are generally consistent with a central dense ISM (hatched absorption region) surrounded by a more diffuse extended wind (gradient). The maximum radius shows the size of the aperture used to generate the mock spectra, while the vertical dashed line shows the column size used to derive the ``true'' simulation column densities. The column size designated by the vertical dashed line lies on the apparent transition between the central galaxy and the wind, ensuring the encapsulation of the majority of the star particles within the absorption region. The $N_{\text{PCM}}(r)$ profile however, remains constant and does not appear to encode any geometric information of the galaxy or wind.
\label{fig:N_r}}
\end{figure}

Here we explore the outcomes of varying the aperture size on inferred column density measurements. Figure~\ref{fig:N_r} shows \SiII{} column density measurements taken from the simulation and estimated from the PCM as a function of aperture radius for a representative sightline. As previously introduced, our fiducial aperture radius for our $N_{\mathrm{sim}}$ columns is 1kpc (dashed vertical line in the figure) and $\sim4$kpc for the mock spectra used to estimate $N_{\mathrm{PCM}}$. First, it is apparent by the near constant $N_{\mathrm{PCM}}(r)$ profile that the choice of aperture size used to estimate the PCM column density is negligble for large enough radii. As long as the bulk of the stellar continuum is contained within the aperture, and additional absorption from a background obect (e.g. galaxy, AGN) isn't present, the absorption lines in the spectrum produce the same PCM column density estimate. Although it has been pointed out in \citet{Gazagnes2023} that reduced aperture size can affect the equivalent width and absorption shape of a line, these changes applied to each transition of the same ion appear to be negligible, producing the same inferred optical depth and covering fraction from the PCM fit (this suggests that absorption line infilling is minimized in the simulation, which would be the case if multiple scattering events in dense gas transfer most of the re-emitted photons to the fluorescent channels). Therefore, the inferred value of $N_{\mathrm{PCM}}$ at our fiducial aperture of radius $\sim4$ kpc is the same value if we were to match the chosen column radius of 1 kpc used to measure $N_{\mathrm{sim}}$.  

Second, the $N_{\mathrm{sim}}(r)$ profiles show a clear change in slope at $r\sim 1$ kpc (the size of our chosen column). These profile shapes are generally consistent with a spherical galaxy geometry of a central dense ISM surrounded by a more diffuse extended wind as shown in the diagram in the left panel of Figure~\ref{fig:N_r}. The largest column density sightline at $r\sim0$ kpc pierces the center of the region of absorption where the stellar continuum lies. The average column density value drops as the size of the column matches the extent of the star forming region and many sightlines pierce tangentially through the central ISM sphere. The slope changes as more of the diffuse outflow is included in the average column density measurement; the average value $N_{\mathrm{sim}}$ tending to zero as $r$ tends to infinity. Therefore, our column radius size of $r\sim 1$kpc to measure $N_{\mathrm{sim}}$ is chosen to contain a majority of the absorbing region in front of the stellar continuum while reducing the effect of skewing the average from large impact parameter sightlines outside the absorption region.

\bibliography{Outflows}{}
\bibliographystyle{aasjournal}



\end{document}